\documentclass{aa}  

\usepackage{graphicx}
\usepackage{natbib}
\usepackage{color} 
\usepackage[varg]{txfonts}
\usepackage[english]{babel}
      \def\new#1 {{\bf #1 }}
      \def\cut#1 {\sout{#1} }

\def\kms {$\mathrm{km\,s^{-1}}$} % km s^-1
\def\degr {\hbox{$^\circ$}}
\def\percc {$\mathrm{cm^{-3}}$} %cm^-3
\def\cmsq  {$\hbox{{\rm cm}}^{-2}$}    %cm-2
\def\cmsqg  {$\hbox{{\rm cm}}^{2}$/g}    %cm-2
\def\Lsol {$\hbox{L}_\odot$}
\def\Msol {$\hbox{M}_\odot$}

\def\AMM {$\mathrm{NH_3}$} %NH3
\def\NTD {$\mathrm{N_2D^{+}}$} %N2D+
\def\GO {G10.99$-0.08$} %T_kin   
\def\GT {G28.33$+0.07$}

\def\simgreat{\mathbin{\lower 3pt\hbox
     {$\rlap{\raise 5pt\hbox{$\char'076$}}\mathchar"7218$}}}
\def\simless{\mathbin{\lower 3pt\hbox
     {$\rlap{\raise 5pt\hbox{$\char'074$}}\mathchar"7218$}}}

\begin{document} 

\title{Massive and low-mass protostars in massive ``starless'' cores}
\author{Thushara Pillai\inst{1,2}, Jens Kauffmann\inst{1,3} , Qizhou Zhang\inst{4}, Patricio Sanhueza\inst{5} , Silvia Leurini\inst{1,6} , Ke Wang\inst{7} , T.K.~Sridharan\inst{4}, Carsten K\"onig \inst{1}}

\institute{Max--Planck--Institut f\"ur Radioastronomie, Auf dem
  H\"ugel 69, D--53121 Bonn, Germany\\
\email{tpillai.astro@gmail.com} 
\and
Institute for Astrophysical Research, 725 Commonwealth Ave, Boston
University, Boston, MA 02215, USA
\and
Haystack Observatory, Massachusetts Institute of Technology, 99 Millstone Road, Westford, MA 01886, USA
\and 
Harvard-Smithsonian Center for Astrophysics, 60 Garden Street, Cambridge, MA 02138, USA
\and
National Astronomical Observatory of Japan, National Institutes of Natural Sciences, 2-21-1 Osawa, Mitaka, Tokyo 181-8588, Japan
\and
INAF - Osservatorio Astronomico di Cagliari, via della Scienza 5, 09047, Selargius (CA), Italy
\and
European Southern Observatory, Karl-Schwarzschild-Str. 2, D-85748 Garching bei M\"unchen, Germany  
}

\date{received XXX; accepted XXX}

% \abstract{}{}{}{}{} 
% 5 {} token are mandatory
 
\abstract{The infrared dark clouds (IRDCs) G11.11$-$0.12 and G28.34$+$0.06 are two of the best-studied IRDCs in our Galaxy. These two clouds host clumps at different stages of evolution, including a massive dense clump in both clouds that is dark even at 70 and 100$\mu$m. Such seemingly quiescent massive dense clumps have been speculated to harbor cores that are precursors of high-mass stars and clusters. We observed these two ``prestellar'' regions at 1mm with the Submillimeter Array (SMA) with the aim of characterizing the nature of such cores. We show that the clumps fragment into several low- to high-mass cores within the filamentary structure of the enveloping cloud. However, while the overall physical properties of the clump may indicate a starless phase, we find that both regions host multiple outflows. The most massive core though 70 $\mu$m dark in both clumps is clearly associated with compact outflows.  Such low-luminosity, massive cores are potentially the earliest stage in the evolution of a massive protostar. We also identify several outflow features distributed in the large environment around the most massive core. We infer that these outflows are being powered by young, low-mass protostars whose core mass is below our detection limit.  These findings suggest that low-mass protostars have already formed or are coevally formed at the earliest phase of high-mass star formation.}
\keywords{Stars: formation -- ISM: clouds -- ISM: jets and outflows}
\titlerunning{Low-mass YSOs in IRDCs}
\authorrunning{Pillai et al.}
\maketitle

%---------------------------------------------

\section{Introduction}

An outstanding puzzle in the formation of high-mass stars is their
earliest phase. What are the initial conditions of high-mass star formation? A frequently invoked theoretical concept is the turbulent core model \citep{mckee2003:turbulence} that predicts the onset of star formation from a monolithic, high-mass, starless dense core supported by turbulence. However,
despite the technical advances in observing techniques and large
surveys spanning the relevant energy spectrum, high-resolution ($<0.1$\,pc) observations have  not been
successful in confirming a signgificant population of such cores due to
(1) their possible absence in massive prestellar clumps
\citep{sanhueza2017} and (2) clear association with molecular
outflows \citep{bontemps2010:cygX, pillai2011a,
  wang2011:g28,lu2015,tan2016a}.  We note, however, that there is only
a handful of interferometric studies on potential sites to find
high-mass prestellar cores and  Atacama Large
  Millimeter/submillimeter Array (ALMA) will play a key role in the near future. 

The identification of such high-mass starless cores is a daunting task in the
heavily clustered, typically distant ($>$ 1 kpc), high-mass star-forming regions. One of the main observational criteria for lack of star formation adopted by
observers is the absence of an infrared (IR) point source at
24 or 70\,$\mu$m. Spitzer and particularly Herschel Space Observatory have been
crucial in distinguishing the starless cores from the more evolved
protostellar cores. However, limited linear resolution at kiloparsec distances combined with high extinction can significantly confuse the interpretation of IR emission (or lack thereof) in high-mass star-forming regions . Therefore, such studies have been combined with astrochemical indicators of very
cold ($<20$ K) conditions such as a high deuterium fractionation and
depletion \citep{chen2010:n2d,fontani2011,pillai2011a,tan2013:hmsc,kong2016}. However, how reliable are such
indirect indicators?  Can high extinction hide very young and low-luminosity protostars within such seemingly starless clumps? If so,
this questions the existence of a high-mass starless stage in
high-mass star formation.  Are the existing cases of high-mass starless cores really
starless?

In this work, we report Submillimeter Array (SMA) 230 GHz continuum and CO
2-1 line observations towards two high-mass 70 $\mu$m dark clumps,
\object{\GO} and \object{\GT}. These clumps are embedded in two of the most well-studied infrared dark clouds (IRDCs), G11.11$-$0.12 at a distance of 3.6
kpc \citep{pillai2006a:g11,henning2010:g11,kainulainen2013,wang2016}
and G28.34$+$0.06 at a distance of 4.8 kpc
\citep{wang2008,butler2012,ragan2012:epos}. These IRDCs host clumps at
different stages of evolution
\citep{zhang09_g28,henning2010:g11,wang2011:g28,
  ragan2012:epos,wang2011:g28,wang2014,zhang2015} and our targets are
the youngest  massive dense clumps in both clouds that are dark  even
at 70 and 100 $\mu$m. At a distance of $< 5$\,kpc, as shown below, we
estimate masses $> 500$\,\Msol\ within 0.6 pc (radius) of the peak
dust continuum emission, with total bolometric luminosity $<
24$\,\Lsol. The dark cloud core \GT\ has been recently extensively studied at  
  millimeter (mm) wavelengths at high resolution, revealing two compact, collimated outflows from the brightest continuum object, which also shows evidence for significant CO depletion and deuteration in \NTD, all indicators of a very young core \citep{chen2010:n2d,feng2016b,tan2016,kong2018:g28}. No high-resolution observations of \GO\ have been reported yet. Here, we present SMA high-resolution
observations of \GO\ and \GT\ focusing on the outflow properties based on CO 2-1 both towards and in the larger vicinity (within 0.6\,pc radius) of the densest clumps.  
%--------------------------------------------------------------------
\section{Observations} 

The SMA\footnote{The
Submillimeter Array is a joint project between the Smithsonian Astrophysical
Observatory and the Academia Sinica Institute of Astronomy and Astrophysics, and
is funded by the Smithsonian Institution and the Academia Sinica.}  observations were made as part of a SMA mini-survey of
high-mass starless core candidates at 230\,GHz. The SMA
 observations were made with all eight antennae in the most compact
configuration (sub-compact) in two tracks in September 2008. The
observations were done in track sharing mode with several sources per track. The
230~GHz receiver was tuned to the CO (2--1) line in the spectral band s13 of the
 upper side band (USB). We used a non-uniform correlator configuration
to provide a higher spectral resolution of $\sim0.2$~\kms\ for the chunks covering
specific spectral lines including CO 2--1. The rest of the correlator was set to $\sim0.4$~\kms. The
data were taken under very good weather conditions ($<1$mm precipitable water vapor), and the
typical system temperatures were 100~K. The primary gain calibrator was J1733-130,
$< 12$~\degr \ away from both targets. Uranus and 3C454.3 were used as the
flux and bandpass calibrator, respectively. Flux calibration is based on the model of the observed planet and is estimated to be accurate to within
20\%. The primary beam at 230\,GHz is $\sim$ 51\arcsec. The data calibration was
done in \texttt{MIR}, an IDL-based calibration package, and then exported to
MIRIAD and CASA for further data reduction, which included continuum
subtraction and imaging. The center of the final image of the mosaic, the
1~$\sigma$ rms noise level achieved in the continuum images, and the
resulting synthesized beam for all observations are listed in
Table~\ref{tbl-1}.  For \GO\ CO 2-1 emission, we combine our
sub-compact configuration data with additional SMA compact configuration data  that were obtained as part of another project
reported by \citet{wang2014}.

We used the Herschel Space Observatory images obtained from the
Herschel Infrared Galactic Plane Survey {Hi-GAL) survey
%ObsIDs: 1342218999,1342218964,1342218965,1342218696,1342218694 and 1342186276; 
\citep{molinari2016} at 70$\mu$m with the Photoconductor Array
Camera and Spectrometer (PACS) instrument.  The standard pipeline data (level 2.5) were downloaded from the NASA/IPAC Infrared science archive.\footnote{http://irsa.ipac.caltech.edu/applications/Herschel/.}

\section{Large-  to small-scale properties of the  massive ``starless'' cores \label{largesmall}}

The two clumps we targeted with SMA are embedded in two of the
  most prominent IRDCs.  Figure~\ref{fig:irdcview}  shows the large-scale, three-color view of the two IRDCs as seen with Spitzer
  \citep{churchwell2009, carey2009}.  It shows the extended
  filamentary extinction features  \citep{butler2009, kainulainen2013}
  and the main star formation sites embedded within them.   Herschel
  (70\,$\mu$m) and APEX  Large APEX BOlometer CAmera (LABOCA) (870\,$\mu$m,
\citealt{schuller2009}) images of the two sources are also presented
in Fig.~\ref{fig:irdcview}.

The brightest mm source (source P1 in \citealt{carey2000:irdc}, see Fig.~\ref{fig:irdcview}) within the
  northern part of the IRDC G11.11-0.12 is a high-mass protostar
  \citep{pillai2006a:g11,gomez2011,henning2010:g11,wang2014}, while the
  southern mm clump we are interested in appears largely
  quiescent.  The focus of this study is this secondary dust emission source (encircled
region in the left panel of Fig.~\ref{fig:irdcview}).  It is clearly associated with a compact absorption feature at 70 and even at
100\,$\mu$m  (see Fig.~B.9 of \citet{ragan2012:epos}). We name this source \GO.
No point sources are observed in the near or mid-infrared with either
GLIMPSE \citep{churchwell2009} or MIPS Galactic Plane
Survey (MIPSGAL).  \citep{carey2009}  as well. 
 The nearest protostellar candidate reported in this
  region is a 25\,\Lsol (5 \Msol) protostar  outside our SMA field of view (FOV) and at a distance of $\sim$
  34\,arcsecs from our central pointing \citep{henning2010:g11}. Single dish \AMM\ observations have shown that this
  region is unusually cold (T$\sim 12$\,K) with a peak H$_2$ column
  density of  2$\times 10^{22}$\,\cmsq\ and total mass of $\sim
    1580$\,\Msol\ \citep{pillai2006b:nh3,urquhart2015}. No high-resolution observations have been discussed in the literature to date.

The IRDC G28.34+0.06 is more complex, with the brightest mm clump in IRDC
  G28.34+0.06 hosting a prominent  IRAS source 18402-0403 (source P2
  in \citealt{carey2000:irdc}), while the next bright mm clump (source P1
  following  \citealt{carey2000:irdc}, Fig.~\ref{fig:irdcview}) hosts a young protocluster
  driving multiple collimated outflows
  \citep{wang2011:g28,zhang09_g28,zhang2015}.  Owing to its large mass reservoir and
  lack of obvious star formation indicators at IR wavelengths, the
  less prominent mm clump we are interested in (encircled
region in the right panel of Fig.~\ref{fig:irdcview}) has been studied in
  detail recently as a potential high-mass starless clump candidate
  \citep{tan2013:hmsc, kong2018:g28}. We name this source \GT. Similar
  to \GO,\, no point source has been reported in  the near or
  mid-infrared with either GLIMPSE \citep{churchwell2009} or MIPSGAL
  \citep{carey2009}. \GT\ is also dark at  70 and even at
100\,$\mu$m (see Fig.~B.22 of \citet{ragan2012:epos}). More recently,  bipolar outflow emission
towards the most massive core has been revealed in \GT\
(\citealt{feng2016b,tan2016a,kong2018:g28}). 

\begin{figure*}
\begin{tabular}{cc}
\includegraphics[width=0.45\textwidth,angle=0]{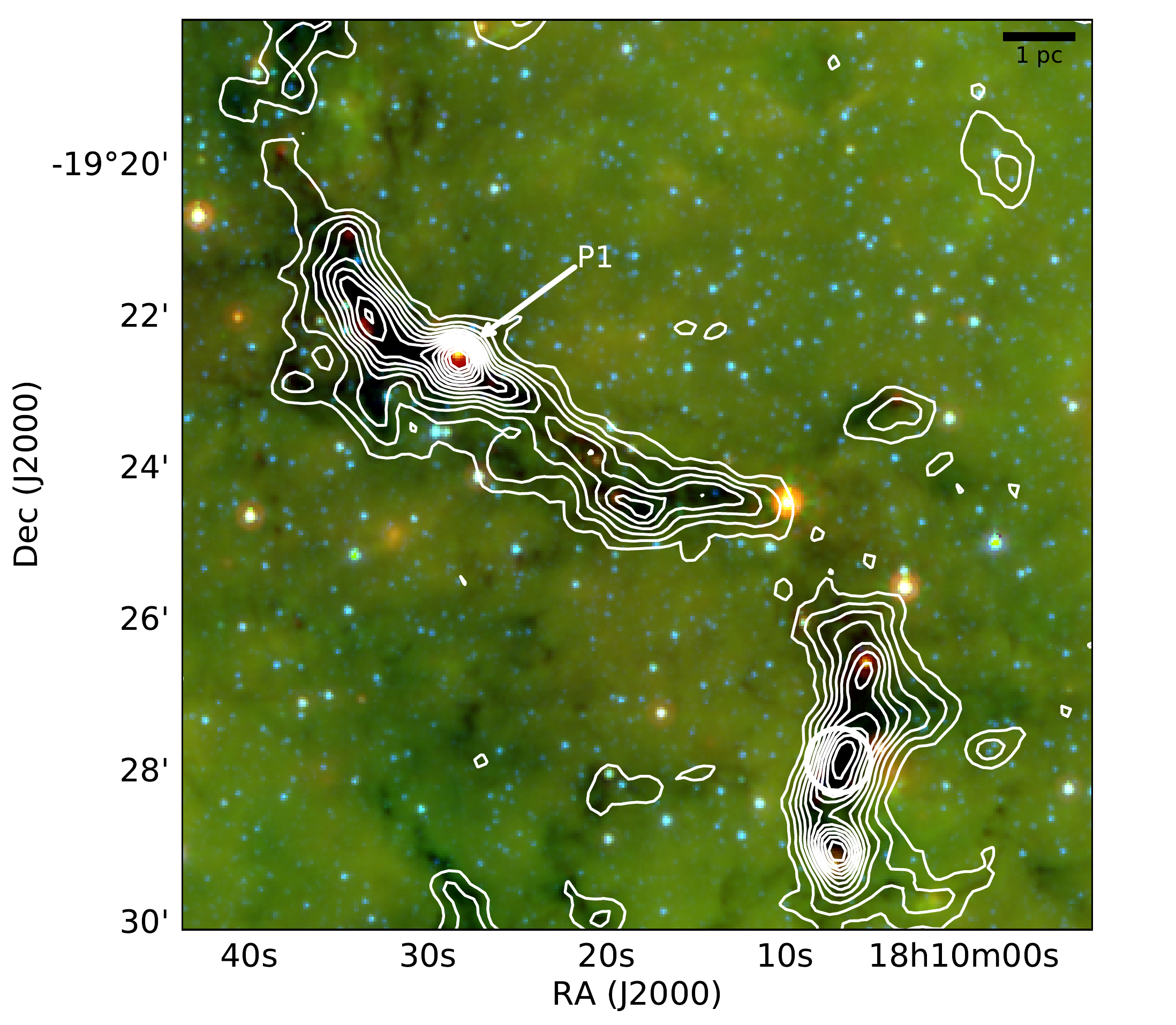} &
 \includegraphics[width=0.45\textwidth,angle=0]{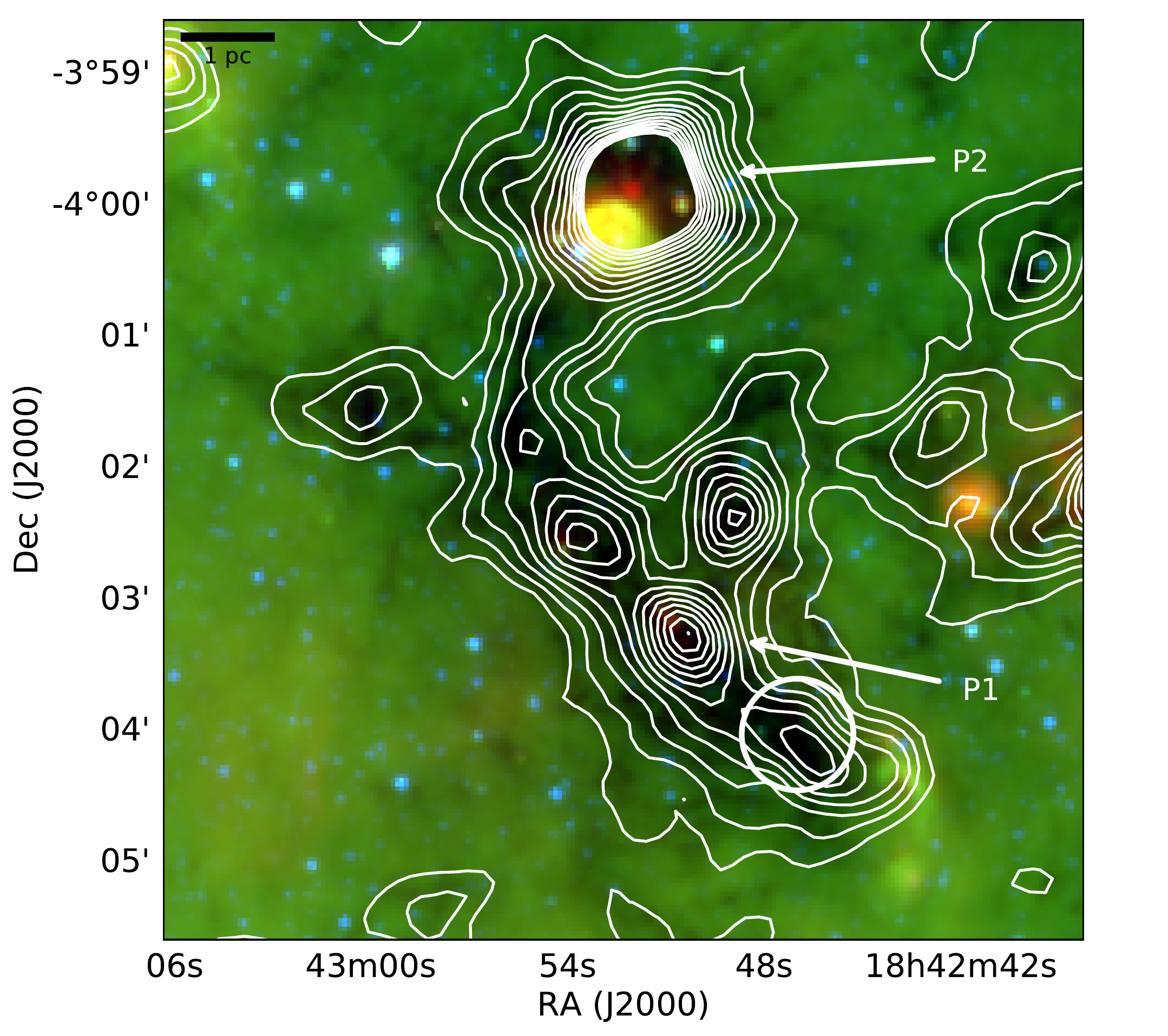} \\
\includegraphics[width=0.45\textwidth,angle=0]{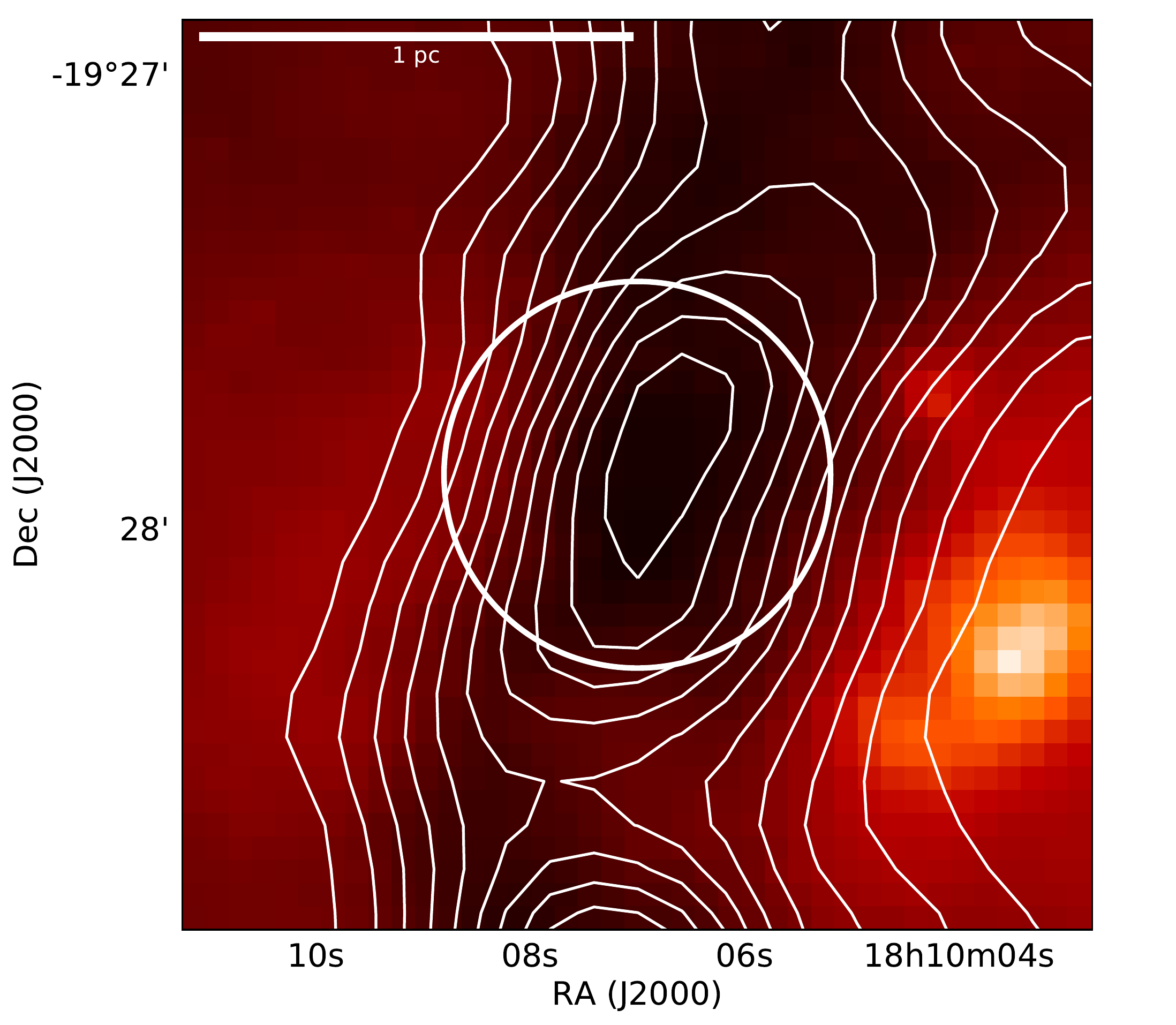}
  & \includegraphics[width=0.45\textwidth,angle=0]{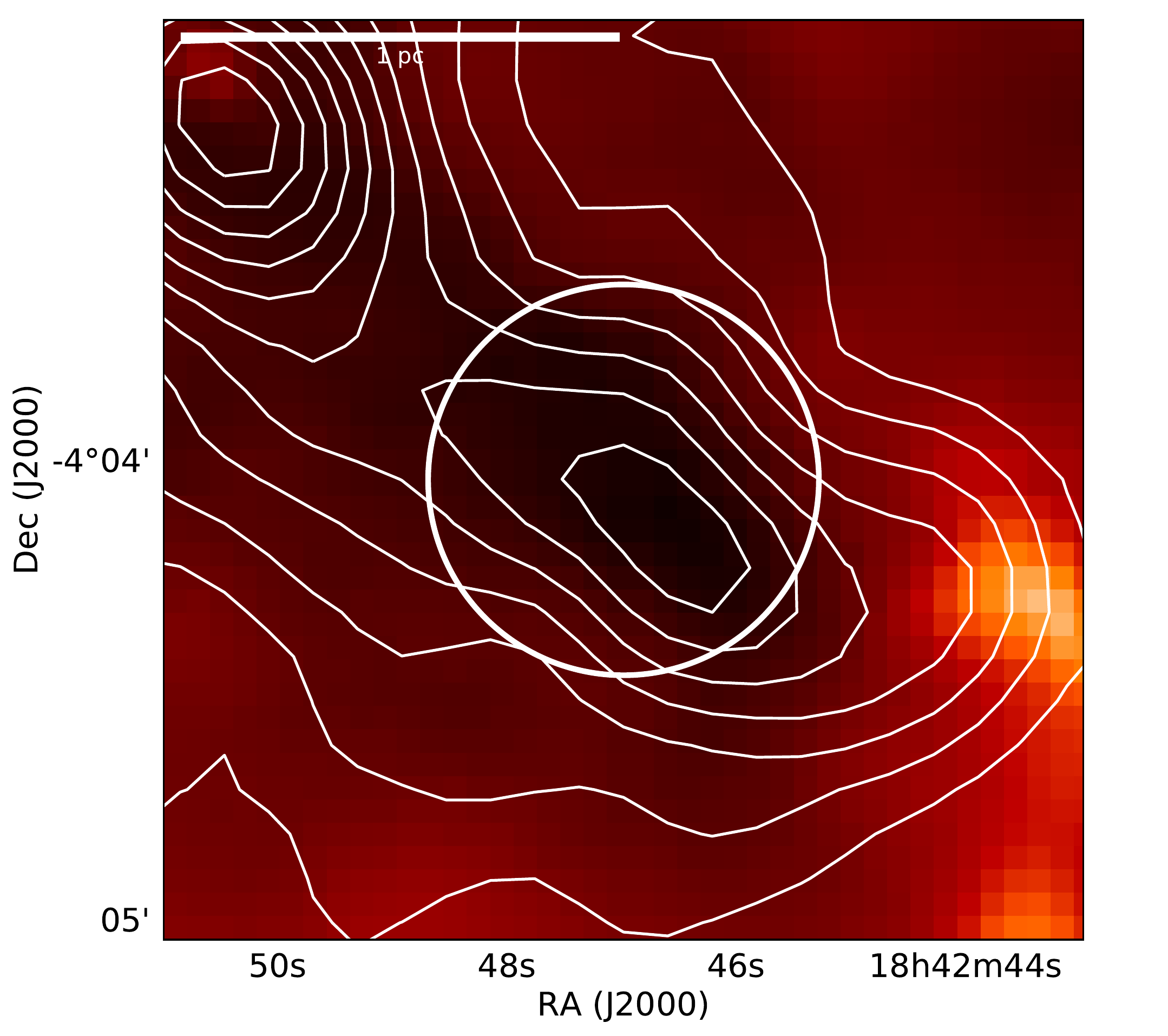}
\end{tabular}
%  \centering
\caption{Top panel: Large-scale, three-color view of the IRDCs
  G11.11-0.12 (left) and G28.34+0.06 (right) obtained with Spitzer IRAC
  and MIPS data with 3.6\, $\mu$m (blue), 8.0\, $\mu$m  (green),
  24.0\, $\mu$m (red) from the GLIMPSE and MIPSGAL survey \citep{churchwell2009,carey2009}. 
Bottom panel:  Herschel
  PACS 70\,$\mu$m emission towards our targets \GO\ (left) and
  \GT\ (right; \citealt{molinari2016}). The contour levels correspond to ATLASGAL
  870 $\mu$m emission (0.1 to 1.3 in steps of 0.1 mJy/beam; 
  \citealt{schuller2009}).   The starless core regions mapped with the
  SMA  are shown as circles. \label{fig:irdcview}}
\end{figure*}

Both observations and physical models of protostellar emission predict
a direct correlation  between 70$\mu$m flux and the internal
luminosity of the source \citep{dunham2008}. \citet{tan2013:hmsc} show
that absorption at 100\,$\mu$m arises from very cold ($<12$\, K) dust
and is indicative of negligible internal heating. We analyzed the
spectral energy distribution (SED) using all Herschel PACS and
Spectral and Photometric Imaging Receiver (SPIRE) bands and LABOCA data. A detailed explanation of the SED fitting is given by \citet{konig2017}. SEDs for \GO\ and \GT\ are
shown in Fig.~\ref{appsed}. The best-fit SED provides a good match to the observed flux
at all wavelengths within the flux uncertainty. It corresponds to a
bolometric luminosity (temperature) of 10 \Lsol\ (9\,K) and $\sim 24$ \Lsol\ (11\,K) for \GO\ and
\GT,\  respectively. These modest luminosities do not imply
the presence of embedded sources: it is merely the luminosity
generated by dust emission from a large mass of cold gas, as also
indicated by the low dust temperatures derived as part of the
gray-body fit. We computed the gas mass for the above-mentioned temperature adopting a dust opacity at 870\,$\mu$m of 0.021~\cmsqg\ for thin ice mantles that have coagulated at a density of 10$^6$\,\percc\ for 10$^5$ yr \citep{ossenkopf1994:opacities}.  The
total mass measured within the SMA field of view (FOV) of 51\,\arcsec\ with single dish (APEX) is 1183 and 960 \,\Msol\ for
\GO\  and \GT,\ respectively. For a radius of  0.44\,pc (\GO) and 0.59\,pc
(\GT), these values exceed the empirical
mass-size threshold for high-mass star formation by a factor $>3$
\citep{kauffmann2010c}.\footnote{As prescribed in \citet{kauffmann2010c}, we scale
  down the submillimeter  opacity by a factor 1.5 when computing the mass-size excess.}
Thus, the SEDs and submillimeter emission for both sources are consistent with
a cold, massive phase with little evidence of ongoing high-mass star formation.

We also used the Spitzer and Herschel non-detections to
constrain the luminosity of the possible population of embedded protostars. The protostellar population is both heavily embedded (for the brightest continuum cores) and/or of low luminosity (distributed population of cores), below the detection limit of infrared observations up to 24\,$\mu$m flux. We can, however, use the information on 70\,$\mu$m flux density to constrain the protostellar luminosity of our targets. Dust radiative models of protostellar emission as well as empirical results from low-mass protostars have shown a strong correlation between 70\,$\mu$m flux and internal luminosity of a protostar. In particular, \citet{dunham2008} show that the internal luminosity of a dense core is related to its flux density at 70\,$\mu$m wavelength, $S_{\nu}(70~\rm{}\mu{}m)$, by (reformulated here for a distance 'd' in kpc)
\begin{equation}
L_{\rm{}int} = 2.4 \, L_{\sun} \cdot
  \left( \frac{S_{\nu}(70~\rm{}\mu{}m)}{\rm{}Jy} \right)^{0.94} \cdot
  \left( \frac{d}{\rm{}kpc} \right)^{1.88} \, ,
\end{equation}
where the numerical constants follow from Eq.~(2) of \citeauthor{dunham2008}. We use this relation\footnote{Alternatively, one could also base the calculation on Table~2 of \citet{dunham2008}, which is based on a sparse sample of observational data. In that case, $L_{\rm{}int}=2.2\,L_{\sun}\cdot{}(S_{\nu}[70~{\rm{}\mu{}m}]/{\rm{}Jy})^{0.94}\cdot{}(d/{\rm{}kpc})^{1.89}$ at $70~{\rm{}\mu{}m}$ wavelength, and $L_{\rm{}int}=25\,L_{\sun}\cdot{}(S_{\nu}[24~{\rm{}\mu{}m}]/{\rm{}Jy})^{0.72}\cdot{}(d/{\rm{}kpc})^{1.45}$ at $24~{\rm{}\mu{}m}$ wavelength.} to place an upper limit on the total internal luminosity of protostars in our targets. Since our sources are dark at 70$\mu$m, we use the 90\% point-source completeness limit  $\sim 0.5$\,Jy for the 70$\mu$m emission reported by the Herschel infrared Galactic Plane Survey, Hi-GAL (Fig. 9 of \citealt{molinari2016}). This provides us with a luminosity upper limit of 14 (24) \Lsol\ for \GO\ (\GT).
This is the predicted  protostellar luminosity resulting
from internal heating and comparable to that derived from our integrated
SED over all wavelengths mentioned above.

At higher spatial resolution with the SMA, the single dish
submillimeter emission in
both sources breaks up into a string of roughly equidistant compact continuum peaks. We
detect compact continuum emission at the position of the 70\,$\mu$m absorption
feature for both targets (see Figure~2).  We identified sub-structures by performing a dendrogram analysis
\citep{rosolowsky2008:dendro} on this dataset to
segment the dust continuum emission. The segmentation starts at a
threshold $\ge 5\,\sigma$ noise level in steps of  $1\,\sigma$  and yields several cores within the 
single dish peak. Adopting a dust opacity at 230 GHz of 0.01~\cmsqg\
\citep{ossenkopf1994:opacities} and a dust temperature of 15\,K, we
also derive total gas mass within the dendrogram structures. Our
3\,$\sigma$ detection threshold corresponds to a mass sensitivity of
$\sim 1$ and $\sim 2$\,\Msol{} / beam for \GO\ and \GT,\ respectively.
The mass range of 1.3 - 33~\Msol\ covers a large range from solar-mass
to high-mass cores  (see Table~\ref{tab:core}). The brightest dust continuum cores (we will refer to it as ``core 1'' in both targets) in both sources exceed  $10$\,\Msol.

The structure of our
continuum emission for \GT\ is consistent with the published ALMA observations of
\citet{tan2013:hmsc}. 
Based on ALMA \NTD\ (3-2)
integrated intensity and dust continuum emission, \citet{tan2013:hmsc} identify two starless core
candidates C1-N and C1-S at 2.3\,arcsec resolution. Their follow-up
ALMA observations at approximately two times better angular resolution reveal that C1-S core
fragments into C1-Sa and C1-Sb \citep{tan2016a}.  Two more dust continuum
cores, C1a and C1b, located away from C1-S and C1-N cores are identified
in this data (see Fig.~\ref{fig:smacont}) .  All four cores are found to drive outflows and are thus
considered protostellar in nature.  Using the mm fluxes given in their
Table~1 and using the same dust opacity and temperature mentioned
above for our SMA observations , we estimate that these cores have masses
of $\sim$ 13 (C1-Sa), 1.8
(C1-Sb), 2.3 (C1a), and 2.3 (C1b)\,\Msol. While our SMA image is
centered roughly on C1-N, the Tan et al.  ALMA pointing center is closer
to C1-S and the SMA 1.3mm primary beam is twice as large as the ALMA
primary beam. Therefore, beyond C1-N towards the north, we should not
expect any overlap in structures. Within the region of overlap,
the position of C1-N  and C1-S coincides well with our observations and C1-N
remains  unfragmented even at the higher resolution of ALMA.  While we
detect some extended emission towards C1-a and C1-b,  at the 
resolution and sensitivity, our dendrogram analysis does not pick
these up as distinct cores.

\section{Outflow search  \label{sec:oflow}}
Our primary goal is to understand the  starless versus protostellar nature of the cores in these apparently ``starless'' regions. Since molecular outflows are one of the most unambiguous tracers of protostellar emission, we search for high-velocity CO 2-1 emission in our high-resolution SMA data.  The outflow emission would be traced in the high velocity wings of CO. 

We  used our previously published  \AMM\ \citep{pillai2006b:nh3} data
to confine the extent of dense gas emission at the systemic
velocity. The systemic velocities are 29.2$\pm3$ and 78.4$\pm
3$\,\kms\ for \GO\ and \GT,\ respectively. Outside of this range, we
define the high-velocity emission from CO. However, our data reveal
the  presence of complex extended CO emission in this high-velocity
range. This emission is uncorrelated with the cloud structure and has
velocities as high as $\sim \pm 50$\,\kms\ around the systemic
velocity. This is exemplified in the integrated spectrum obtained on
the brightest continuum sources (see Fig~\ref{appcospec}). We note
  that the negative spectral features in Fig.~\ref{appcospec} arise
  from missing short spacing information from diffuse emission in an
  interferometer image. A significant fraction of emission features do
  not show any clear collimated structure that could be associated
  with an outflow component.  The presence of such complex emission is
  not unexpected given that the two clouds are located well within the
  inner Galactic Plane $|b|<0.2$ degrees and therefore could be
  potentially contaminated by unrelated interstellar medium (ISM) features along their line of sight. 
A wide range in velocities between -10 and +50\, \kms\ would be consistent with galactic rotation in the direction of both clouds (see Fig. 3, \citealt{dame2001}) and can contaminate $^{12} \rm{CO}$ data from unrelated ISM features along the cloud line of sight.  Such ISM structures may be dense ($\rho > 10^4$\,\percc) or diffuse. Dense structures, however, would also be detected in dense gas tracers such as \AMM. There is no indication for such a component outside of the systemic velocity range discussed above. Diffuse emission along the line of sight with little sub-structure would be mostly filtered out by the interferometer. However, the SMA in its sub-compact configuration is
sensitive to structures as extended as $\sim 0.5-0.7$\,pc at $3.6-4.8$ kpc (for a minimum baseline length of $\sim 9.5$\,m). 

In order to extract only the outflow related component, we do the following.
First, we use a 1 dimensional non-LTE line radiative transfer model RADEX
\citep{vandertak2007} to calculate the expected brightness temperature of a
diffuse ($\ge 1 A_V$) cloud component along the line of sight. Using the
standard $^{12} \rm{C}$/$^{13} \rm{C}$ ratio (77) and $^{12} \rm{CO}$ abundance
($2\times 10^{-4}$), density of 1000~\percc\ . and temperature of 15~K, we derive
a brightness temperature $> 0.5$\,K. We extract $5\times 5 $ arcmin$^2$ maps of
$^{13} \rm{CO}$ 2-1 map from APEX \citep{schuller2017} and 1-0 map from
Five College Radio Astronomical Observatory (FCRAO) \citep{jackson2006} for \GO\ and \GT,\ respectively. We then identify
velocities in the channel maps associated with all structures extended with respect to the beam
above the 0.5\,K threshold. These velocity ranges are explicitly excluded when
creating moment maps in the $^{12} \rm{CO}$ SMA data.  Due to the
  expanding motions, outflows would exhibit sharp velocity gradients
  when visualized in the position-position-velocity (PPV)
  space, while intervening clouds would have a  sheet-like
  distribution.  Therefore,  as a final check, we use a 3 dimensional volume rendering  data visualization software package Glue\footnote{http://glueviz.org.} to explore structures in PPV space and only consider structures associated
with outflows. Additionally, for the two clouds, we also flag out the
shortest baselines  ($<10 \,\rm{k\lambda}$ or 13m)  to further filter
out large-scale structures. A $uv$ distance of $10 \,\rm{k\lambda}$ is
merely the best compromise that we found between retaining outflow
emission while removing large-scale features and does not have any
physical significance. We note that this process would also remove some
outflow emission (and thus underestimate outflow properties). Due to
the nearer distance (sensitive to more extended emission for a given
beam) and more severe line of sight contamination, we apply this
filtering for the whole velocity range for \GO. For \GT\, only the
velocity range 84-88\,\kms\ (more than 6\,\kms\ red-shifted from the
cloud  local standard of rest (LSR) velocity) is subjected to the filtering constraint since
there is  significant emission from outflows as well as ambient gas
only in this range. The resulting $^{12} \rm{CO}$ structures are shown
in the right panels of Fig.~\ref{fig:smacont}.  This emission
is complex and not always bipolar  presumably because the outflows we detect at our resolution are unresolved and likely to be a combination of outflows from unresolved cluster members. Therefore, defining specific outflow features has limited meaning and we shall revisit this with high-resolution, high-sensitivity ALMA data (Sanhueza et al. in prep.). 

Having made an
extensive effort to exclude  emission unassociated with the cloud, we conclude that the high-velocity
emission shown in Fig.~\ref{fig:smacont} is very likely associated
with outflows. Thus, we have our final maps that allow us to derive
combined outflow properties of the system such as mass, momentum, and
energy. We do not calculate dynamical timescales (and therefore
mechanical force, momentum, and energy injection rates) for these
outflows. Derived outflow properties are listed in
Table~\ref{tab:of}. It is important to note that we ignore velocities
close to the ambient velocity, adopt an CO excitation temperature at
the lower end of the usual range of 10-50\,K, assume that $^{12}
\rm{CO}$ is optically thin, and we do not account for the (unknown)
inclination angle of the outflow and spatial filtering of extended
emission in our observations. When accurately constrained and
considered in the mass calculation, all these effects conspire to
significantly increase the outflow mass by an order of magnitude or
more \citep{dunham2014}. Therefore, our outflow parameters are
strictly very conservative lower limits. In fact, deeper ALMA
observations of the most massive outflows in \GT\ \citep{tan2016a}
find outflow parameters that are an order of magnitude higher than our
values.  For \GT\, we compared in detail our SMA outflow
  detections (our Fig.~2, bottom right panel) with that of the
  higher resolution outflows detected in Tan et al.  (their
  Fig.~1, bottom right panel). It is evident from this comparison that the main outflow emission in our SMA data towards the  unresolved source
  C1-S is actually two outflows driven by C1-Sa and C1-Sb.  Tan et
  al. find the outflow masses ($\sim 1$\,\Msol) and momenta ($\sim
  11$\,\Msol\ \kms) from C1-Sa and C1-Sb to be similar within the
  expected errors. In their more recent work reporting even higher
  resolution observations (0.2\,\arcsec),  \citet{kong2018:g28} report that
  the CO outflow momentum flux of  the lower mass  source C1-Sb is only moderately
smaller and has a larger outflow cavity opening
angle  than that of C1-Sa, suggesting that  C1-Sb is a low- or intermediate-mass protostar at a more
advanced evolutionary phase. Several red and blue-shifted
features outside of these two outflows appear in their map.  The lack
of resolution makes it difficult to be certain, however these would
roughly overlap with  our weak outflow features towards C1-N, C1-a, and
C1-b. The nature (such as mass or momentum) of these outflows are not discussed in Tan et al. or
Kong et al. The majority of our brighter outflow features are located outside
their FOV. At a similar sensitivity level, collimated outflows have been detected with the SMA towards IR bright protostars in the two IRDCs harboring \GO\ and \GT\ \citep{wang2011:g28,wang2014}. These outflows have outflow masses and momenta much higher on average than those towards the IR quiet regions that we are studying in the same cloud. These differences are most likely  due to the differences in their evolutionary stages as also evidenced by the different IR luminosity.  We note that outflow properties are similar to those reported for low-mass protostars in a more evolved high-mass protocluster by \citet{cyganowski2017}.

\begin{figure*}
\begin{tabular}{cc}
 \includegraphics[height=0.54\textwidth,angle=90]{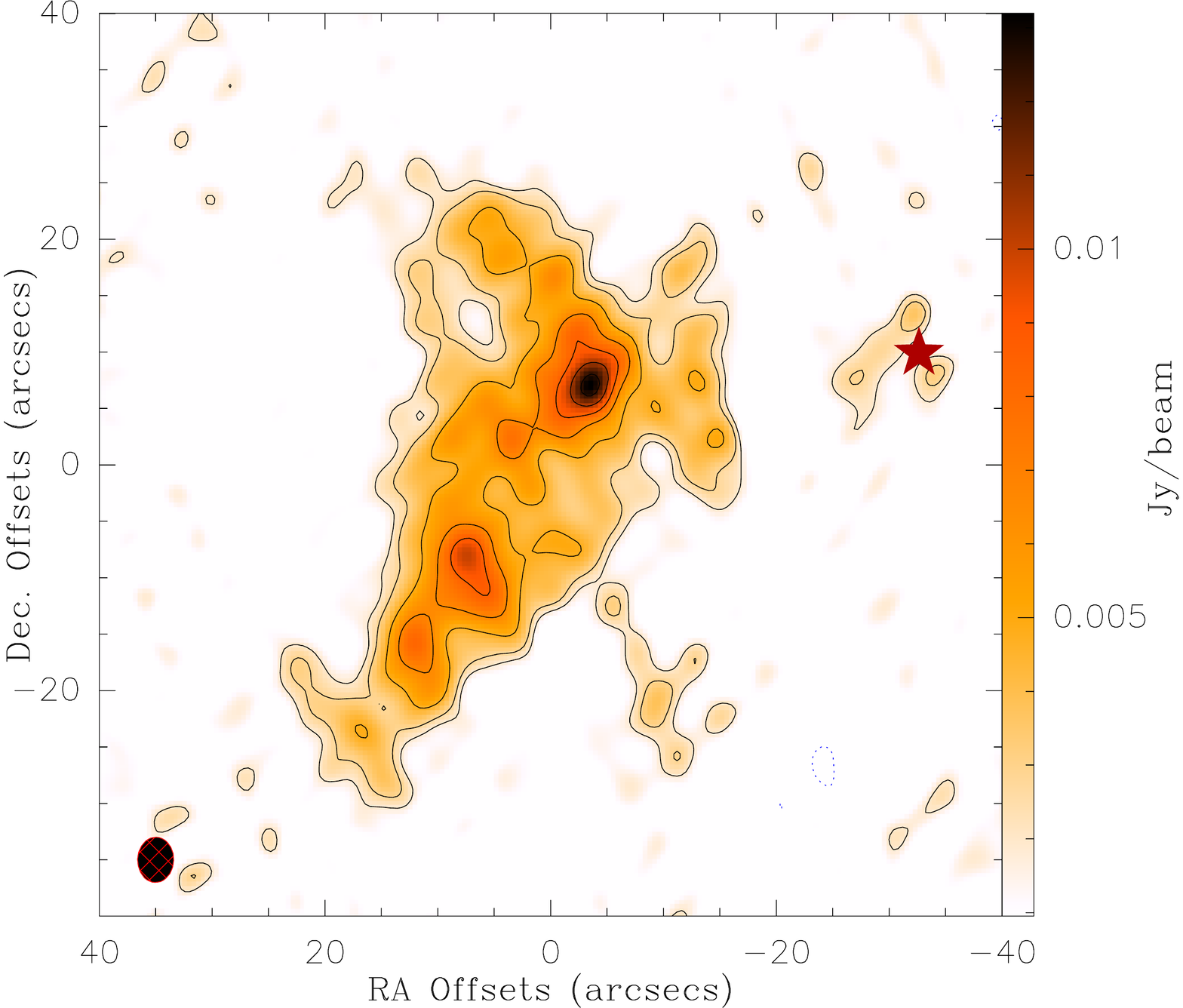}
 &  \includegraphics[height=0.38\textwidth,angle=0]{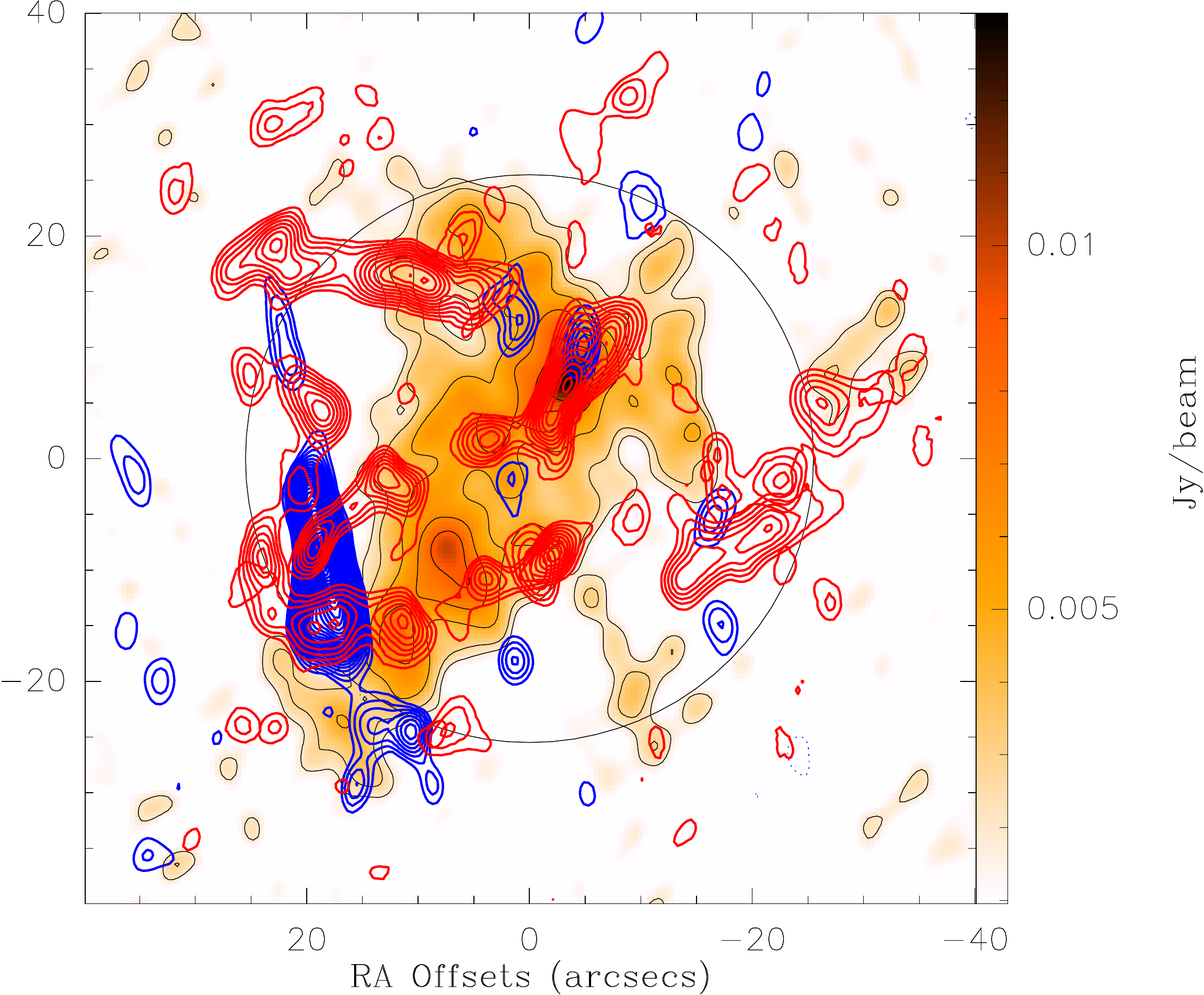} \\
\includegraphics[height=0.54\textwidth,angle=90]{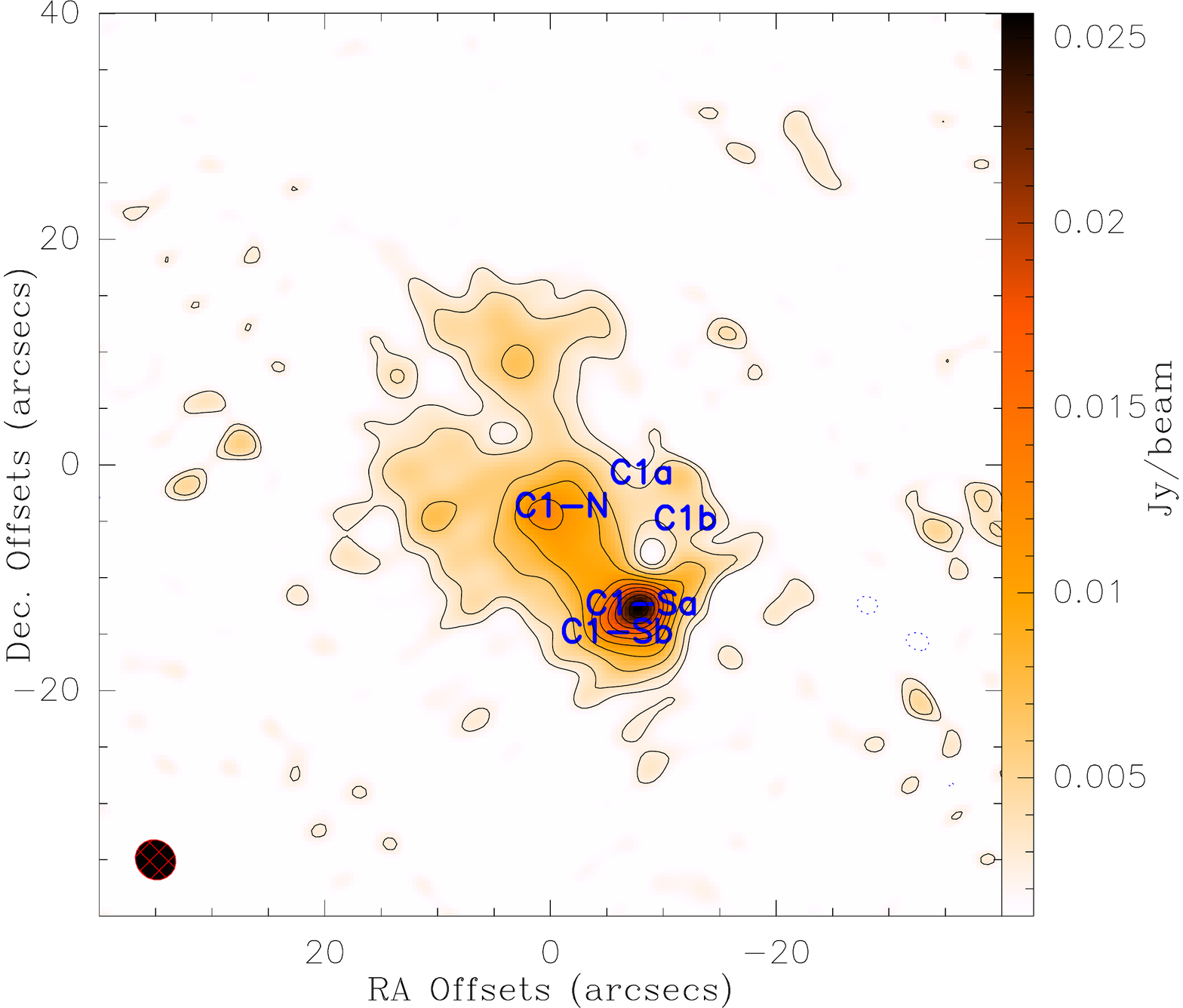} & \includegraphics[height=0.38\textwidth,angle=0]{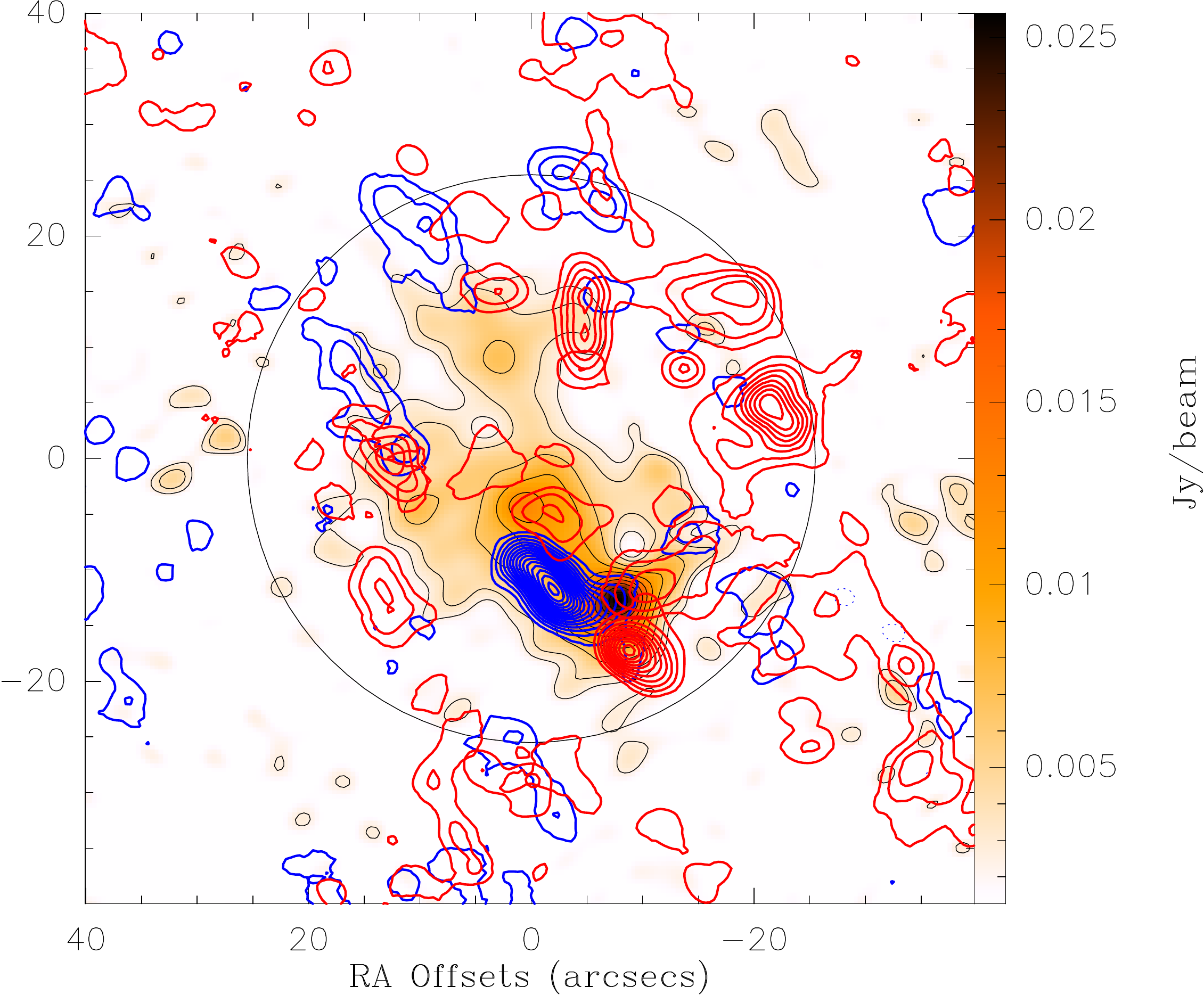}
\end{tabular}
%\centering
\caption{Left panels: SMA 230 GHz thermal dust continuum
  emission towards \GO\  (top) and \GT\  (bottom). The levels start at $2\,\sigma,
  3\,\sigma$, in steps of $2\,\sigma$. Right panels:  SMA CO 2-1 integrated
  intensity emission at high velocities as contours overlaid on SMA
  dust continuum observations towards \GO\  (top) and \GT\  (bottom).  The contours start at $6\,\sigma$, in
  steps of $2\,\sigma$. The circle shows the SMA FOV at 230
 GHz. The ellipse in the lower left corner represents the SMA
 synthesized beam size. The magenta star in the top left panel for
 \GO\ corresponds to a Herschel identified low-mass protostar outside
 the SMA FOV by \citet{henning2010:g11}.}  
\label{fig:smacont}
\end{figure*}

\section{Discussion}

We detect several outflows within $\sim$1 arcmin diameter of the two
targets. This is intriguing given the lack of infrared emission within
the whole clump.

\subsection{Nature of the 24/70\,$\mu$m dark massive core:
    High-mass star progenitors}
In both targets, the brightest continuum core of $>10$\,\Msol\ is
associated with at least one outflow. In \GT,\, high-resolution ALMA
observations have confirmed two distinct outflows. The outflows newly
detected in \GO\ by us validate that both massive apparently starless
cores are in fact not starless.
 At 0.005 pc resolution, the most massive core in the \GT\ region has 30\,
  \Msol\  \citep{kong2018:g28}.   At a roughly similar spatial resolution (0.007 pc), the most
  massive core in nearby low-mass clusters like NGC~1333 has a mass of
  2\Msol\  \citep{plunkett2015}.  Similarly, ALMA higher-resolution data for \GO\ suggest that
  the main core (core 1) shows little fragmentation as also suggested
  by its compactness at SMA resolution (Sanhueza et al. in prep). The
  outflow wings of the most massive core in both sources extend to
  $>50$\,\kms\ (this work, Sanhueza et al. in prep.,
  \citealt{kong2018:g28}), which is typical of high-mass outflows
  \citep{maud2015}. These outflows are also very compact with respect
  to the SMA beam, implying a highly collimated powerful outflow. It
  is clear from the core mass, compactness of the outflow, and its
  extent to very high velocities ($>50$\,\kms) as seen towards massive
  outflows that the most massive cores (core 1) in \GO\ and \GT\ stand
  out in their outflow properties from the rest of the outflow
  features.  These data in summary suggest that high–mass stars in
    an early state of evolution are the drivers of these outflows. However, can a protostar with a massive envelope
  exhibit such a low luminosity of only 10 to 20\,\Lsol\ and still build up over time to a
  massive star?

 To address this question, we compare the observed envelope properties such as mass and
  bolometric luminosity to that predicted by radiative transfer (RT) models
  focused on massive young stellar objects (YSOs). A large RT model grid has been developed
  for SED of massive YSOs by \citet{molinari2008}. Different
  models have been explored  for five different final stellar masses
  from 6.5 to 35\,\Msol.  Comparing our observations to their model
  predictions, the observed low bolometric luminosity of 10-20\,\Lsol\ is to be expected for
massive cores in the first 50,000 years of evolution. Models with
final stellar masses of 6.5, 8, and 13.5\,\Msol\ reach such low
luminosities within a short time of $<50,000$ years (see their Fig.
9 and Table ~7).  According to \citet{molinari2008} , these objects
would be the Class 0 analog in the classical classification established
for low-mass YSOs. The duration of the phase consistent with our
observations rapidly decreases with increasing stellar mass. It thus
seems plausible that the actual duration of the low–luminosity phase
observed by us is well below the upper limit of 50,000 yr. Models of more massive
stars (18\,\Msol\ and higher) are not consistent with our data. 

We also compare our observations with radiative transfer models based
on the turbulent core theory of massive star formation
\citep{mckee2003:turbulence}.  These models include protostellar
evolution in a self-consistent matter
\citep{zhang2014:model, zhang2018:rt}.  Depending on the initial mass
of the turbulent core, (M$_{\rm c}$), and mean clump mass surface
density, $\Sigma_{\rm cl}$ , the final stellar masses in their model
are in the range 20 to 90\,\Msol.  Figure~6 in
\citet{zhang2014:model} follows the evolution of  massive protostellar
cores with different $\Sigma_{\rm cl}$.  For $\Sigma_{\rm cl}=0.316$\,g/cm$^2,$ which is
roughly representative of our clumps, progenitors of massive protostars have luminosities as
low as our observed luminosities in the first 20,000 years of
evolution.  We thus conclude that the embedded stars of low luminosity in our target regions are consistent with being young, high-mass stars. It is impossible to determine the evolutionary status of these sources with absolute certainty. Still, the combination of core mass, strong outflow, and modest limits to the luminosity make young, high–mass stars the most plausible candidates for these sources.

The predicted model luminosities and therefore ages from both these models, however, might be significantly offset
from the actual observed luminosities.  Known as the “protostellar luminosity problem”, the observed
protostellar luminosities of low-mass protostars are approximately ten
times
less luminous than expected from theories of protostellar evolution
\citep{kenyon1990,evans09:sf_rates}.  Out of the many solutions
proposed to explain this discrepancy, the most favored is that of 
an episodic accretion, where an accretion is extremely variable and
the bulk
of the protostellar mass is accreted in rare bursts of high
accretion. Recently, observational evidence
for such accretion events has also been building up for massive stars
\citep{Caratti-O-Garatti2017,hunter2018}. If such accretion events are
indeed the main source of accretion, then the model luminosities
discussed above need to be refined by an order of magnitude or
higher. This would imply that realistic luminosities from models should
be lower and the protostellar lifetimes longer than those predicted
by the current models.

The large-scale mass reservoir ($\sim$\,1000\,\Msol\ in both targets) would
allow the  formation of massive protostars in both sources (see, for
example, \citealt{sanhueza2017}). For \GT,\,  higher resolution
  ALMA observations reveal a protostellar envelope towards the central
  object of between 20 and 50\,\Msol\ \citep{kong2018:g28}.  Additionally, SiO
  2-1 and CO 2-1 outflow properties discussed in  \citet{feng2016b}
  and \citet{tan2016a} respectively,
 support the scenario that the central object is a massive protostellar core that is in a relatively early phase of collapse.  Based on
the models discussed above that are relevant to high-mass protostellar
evolution, we conclude that for final stellar masses between 6.5 and 20\,\Msol,  we
can indeed reproduce the observed low luminosity and protostellar masses for
\GO\ and \GT\ as being associated to YSOs at the earliest stage in the evolution of massive protostars.

 \subsection{Nature of other outflow driving sources without mm continuum}

What we see in the larger environment of the clump away from the
  main continuum core is a complex superposition of outflows.  Since
  these  outflows are not associated with any continuum cores,  it is
  important to compare the combined structure of these outflows with
  nearby, resolved, better understood templates of overlapping
  outflows.  We use the nearby, resolved young cluster region in NGC~1333
as our outflow template. We chose this particular region because the
spatial resolution of these observations match the spatial resolution
of our SMA \GO\ observations ($\sim 0.06$\, pc at 3.6 kpc). Clearly
our region hosts  more massive cores than NGC~1333, however it also
appears to harbor a small, distributed (over $\sim 1$\,pc) population
of young, low-mass YSOs at the same time.

We note that for \GT, in the region where there is spatial overlap between our SMA data and
ALMA data in \citet{tan2016a}, Tan et al. report two low-mass dust continuum
cores of $\sim 2$\,\Msol\ named C1a and C1b (see
Sect.~\ref{largesmall}) that drive outflows and are thus
considered low- or intermediate-mass protostars.  Most of our outflows are outside
this region of overlap for \GT.  

First we investigate the effect of a limited resolution on
the CO observations. We used a published CO 1-0 outflow map
\citep{arce2010} towards the NGC~1333 star-forming region within the
Perseus molecular cloud.  Observations have shown that the NGC~1333
region is forming low- and intermediate-mass stars and is the most
active site of star formation in Perseus \citep{bally2008}. We selected
an active sub-region ``Area II'' following the naming convention in
Arce et al. (see their Fig.~6). Arce et al. detect several outflows
in the region and report the extent of outflow emission in blue and
red lobes. We  downloaded their archived FITS data cubes and
generated integrated intensity map for the average range identified in
their work (see their Fig.~4). Adopting the known distances to
Perseus (250 pc) and \GO\ (3.6 kpc), we scaled the integrated intensity
image to the distance of the \GO\ region. We then extracted a 1 arcmin
region (roughly matching the SMA field of view of $\sim 51$ arcsec)
around the scaled image and performed a visual comparison of the outflow
properties. As shown in Fig.~\ref{fig3}, five outflows from Arce et
al. (Flows 1--5) have been identified in red-shifted emission within the same spatial scale as our SMA data. The total mass within all outflows is 3.5\,\Msol\, while we can only place a lower limit of $\sim 0.05$\,\Msol\ for the outflow mass in \GO. We have not taken into account the effect of interferometer spatial filtering that would filter out the large-scale emission of these outflows and would make them appear more compact and less massive. The number of outflow features, and the general overall morphology of the features identified as potential outflows in our data is therefore broadly consistent with what one might expect based on the resolved emission from a low-mass cluster-forming region like NGC~1333. 

The lack of thermal dust continuum cores is to be understood as well. Our dust
continuum observation is sensitive to $\sim 1-2$\,\Msol\ core. The
sources in NGC~1333, however, have sub-solar masses
(\citealt{plunkett2013}, except for IRAS4A), significantly below our
detection limit. In the mid-Infrared (mid-IR), \citet{dunham2015}
report 125 protostars based on MIPS 24\,$\mu$m band observations. We
used the Spitzer MIPSGAL Legacy Program that imaged the 24$\mu$m inner
galactic plane \citep{carey2009} to search for mid-IR counterparts
driving  outflows in \GO\ and \GT. Specifically, we used the 24$\mu$m
point source catalog based on MIPSGAL \citep{gutermuth2015} to
identify all protostars within the SMA FOV. This search yields only a
single point source in \GO\ and none in \GT. The location of the 24$\mu$m
point source in \GO\ is distinctly offset from all dust continuum
peaks along the main ridge. In order to understand the effect of
distance, we scaled the Dunham et al. point source flux densities to
those at a distance of 3.6 kpc and set a sensitivity cutoff at 1 mJy
similar to that used by \citet{gutermuth2015}. We can then predict
that $\sim 21$ YSOs in NGC~1333 are expected to be detected at 3.6
kpc.  However, the extinction maps towards \GO\ and \GT\
\citep{kainulainen2013, butler2009} reveal that deep within the
galactic plane, even the larger scale
environment of the IRDCs that we are probing experiences high extinction. A 30~mag visual extinction can
easily diminish the flux densities at 24$\mu$m by a factor of five \citep{flaherty2007}, resulting in an equivalent detection threshold of 5\,mJy. This would reveal at most  approximately four protostars at 3.6\,kpc in NGC~1333, consistent with our finding of a single point source in the catalog above a cutoff of 1\,mJy. Therefore, we conclude that the observed complex outflow emission, as well as the lack of
  continuum cores and MIR sources associated with these outflows are
  consistent with a distributed population of low-mass protostars.

\subsection{Cluster formation in IRDCs} An initial search for low-mass
star formation towards a distant high-mass star-forming IRDC revealed
a distributed population of low-mass pre-main sequence stars \citep{foster2014}.  These
were near-IR (K band) observations that could only detect low-mass
protostars forming in tghe inter-clump medium of the IRDC and not
towards the high column density clumps where massive protostars are
embedded. A massive more evolved (IR-bright) clump located within the
same IRDC as \GT\ was followed up with high sensitivity ALMA
observations \citep{zhang2015}. These observations together with a
careful treatment of interferometer effects did not find any evidence
for a distributed population of low-mass sources either in the form of
low-mass cores or outflows close to the high-mass cluster. This led
Zhang et al. to conclude that low-mass stars would form later, while they also found that lower mass cores are forming within the immediate vicinity of higher mass objects. In contrast to their results, a recent ALMA observation of
a more evolved massive star-forming region have found evidence for a
population of low-mass cores in the outer boundaries ($\sim 0.17$\,pc
median value) of the main mass reservoir \citep{cyganowski2017}. The
largest spatial scales probed might play a role in the interpretation
of low-mass protostars. Cyganowski et al. observations were made in
mosaic mode covering a spatial extent of 0.7 pc, while the Zhang et
al. study is a single pointing study. Different initial conditions or
differences in evolutionary phase might also play an important role. Our observations discussed here towards an even earlier phase of massive star formation than studies so far, show potential evidence for low-mass star formation distributed in a larger (0.6~pc) environment around the accreting massive core. This is gathering evidence for a scenario where low-mass stars either form first or form coevally with high-mass stars.

\section{Conclusion}
Clusters play a very important role in the formation of massive stars \citep{lada2003:araa}. However, we are yet to understand the detailed physics of this process. 
Our goal was to understand massive star and cluster formation at the earliest phase by  studying two massive cold clumps seemingly with no evidence of star formation.

\begin{itemize} 
\item We detect high-velocity emission in CO 2-1 towards two 100$\mu$m dark massive clumps that show no other evidence of star formation.
\item  The most massive cores in both sources are driving
  outflows. Their properties are consistent with the youngest phase
  ($<50,000$\,yrs) in evolution of a massive protostar. 
\item Our observations reveal several other potential outflow features distributed across a region of 0.6~pc distinctly separated from the central massive cores. 
\item The lack of  massive continuum cores and the combined outflow properties indicate that the driving sources of these sources are likely of low-mass (solar or sub-solar) origin.
\item Our observations towards such a very early phase of massive star formation indicate that low-mass stars might form first or coevally with massive stars.

\end{itemize}

\begin{acknowledgements}
 We thank the anonymous referee
for a critical and constructive review of this manuscript. T.P. and K.W. acknowledge support from the \emph{Deut\-sche
  For\-schungs\-ge\-mein\-schaft, DFG\/} via the SPP (priority
program) 1573 ‘Physics of the ISM’.  P.S. was financially supported by Grant-in-Aid for Scientific Research (KAKENHI Number 18H01259) of
the Japan Society for the Promotion of Science (JSPS). This publication makes use of
molecular line data from the Boston University-FCRAO Galactic Ring
Survey (GRS). This research has made use of the VizieR catalog
access tool, CDS, Strasbourg, France. 
%TP thanks Felipe Navarette for searching for H2 emission in the near-IR data (UKIRT) . 
\end{acknowledgements}


\bibliographystyle{aa}
\bibliography{bib_astro}

 \clearpage

\begin{appendix} %First appendix
  \section{Dust continuum core properties and noise-levels}
  
 \begin{table*}[h]
\begin{center}
\caption{SMA 230 GHz continuum sensitivity.\label{tbl-1}}
\begin{tabular}{ccccc}
Source   & R.A.  & Dec. & Sensitivity$^a$   & beam$^b$ \\ 
              & (J2000.)        & J2000.             & 1$\sigma$ (mJy)  &  (arcsecs) \\
\hline
\GO\              & 18:10:07.00 & -19:27:52.90    &  1.0   & 4.0$\times$3.2  \\
\GT\                & 18:42:46.98 & -04:04:02.600   & 1.0     & 3.7$\times$3.4 \\
\end{tabular}
\end{center}
$^a$: The maps are dominated by emission and therefore the noise estimate is an upper limit.
$^b$: We report the full-width at half-maximum beam (FWHM) of the synthesized beam
\end{table*}

\begin{figure}
\begin{tabular}{c}
\includegraphics[height=0.2\textheight,angle=0]{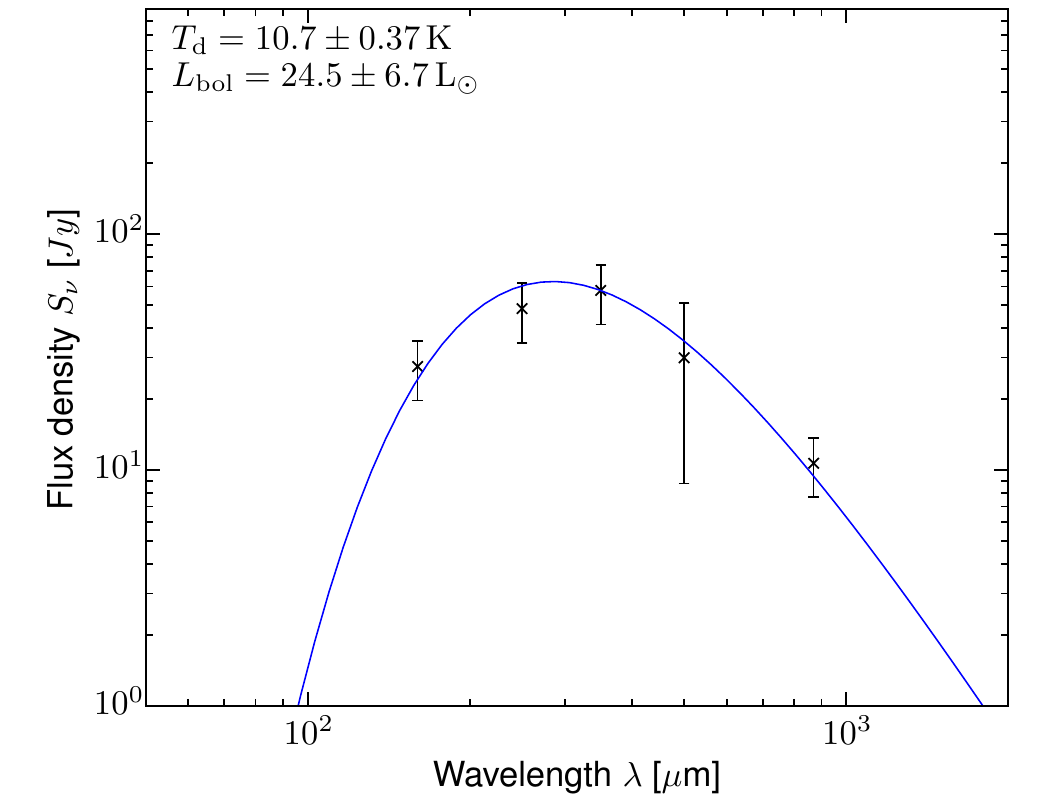}
  \\
\includegraphics[height=0.2\textheight,angle=0]{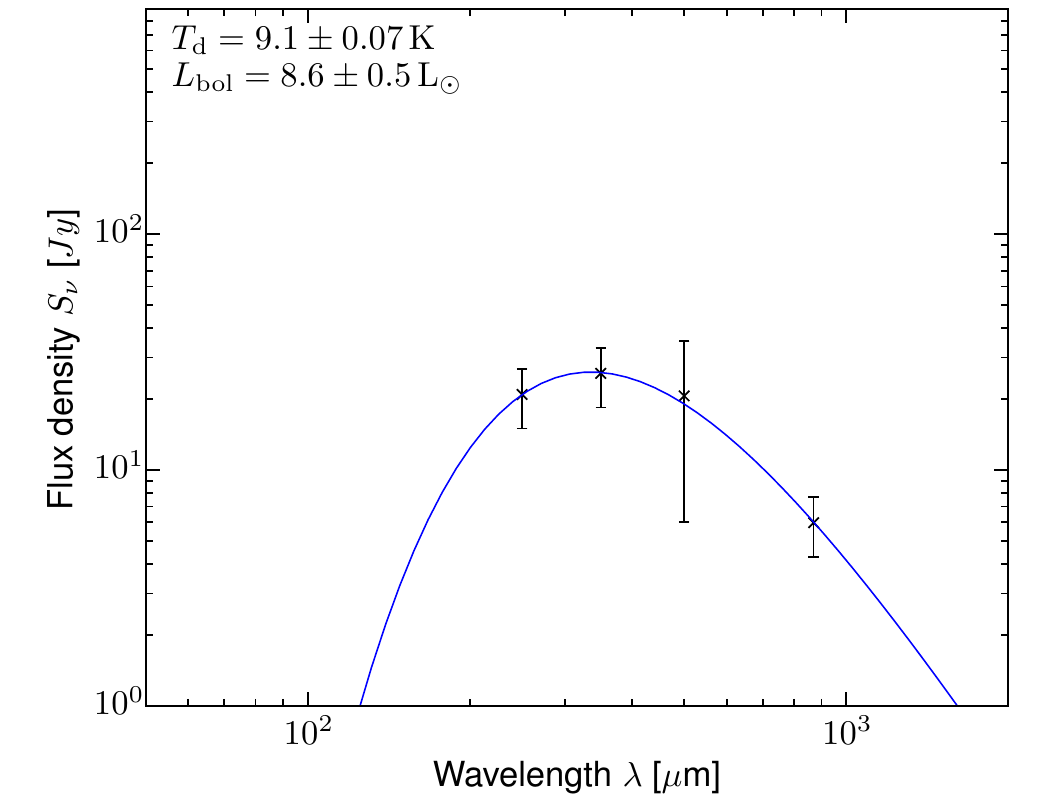}\\
\end{tabular}
\centering
\caption{Spectral energy distribution based on Herschel \citep{molinari2016} and LABOCA data \citep{schuller2009} for \GO\ (top panel) and \GT\ (bottom panel). The data at
  different wavelengths are shown as squares with error bars.  
  The temperature and  bolometric luminosity obtained based on the
  SED fit assuming a dust opacity ($\beta=1.8$) is also shown \citep{konig2017}.}
\label{appsed}
\end{figure}

\section{Outflow search} %Second appendix

\begin{figure}
\begin{tabular}{c}
\includegraphics[height=0.2\textheight,angle=-90]{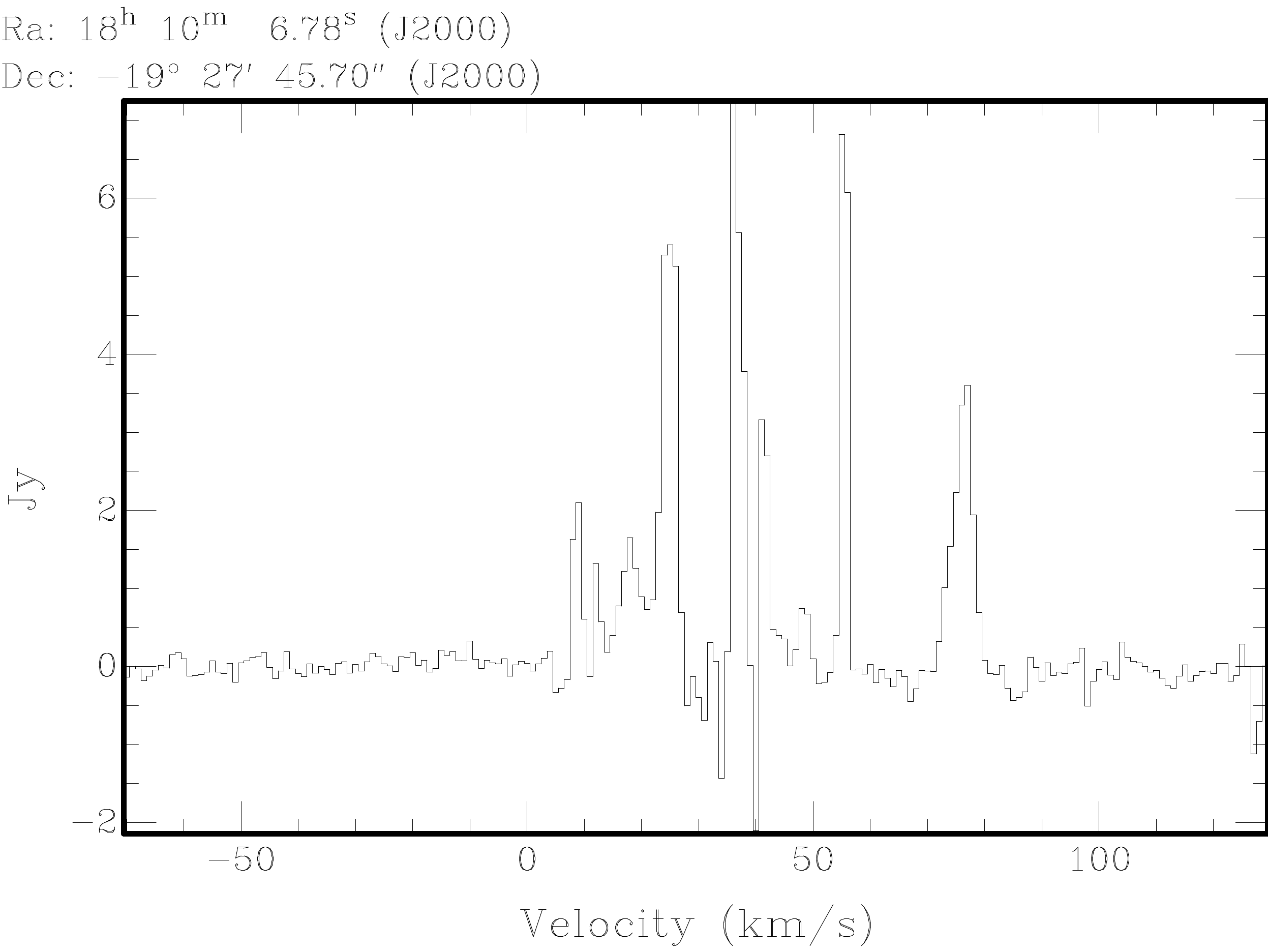}
  \\
\includegraphics[height=0.2\textheight,angle=-90]{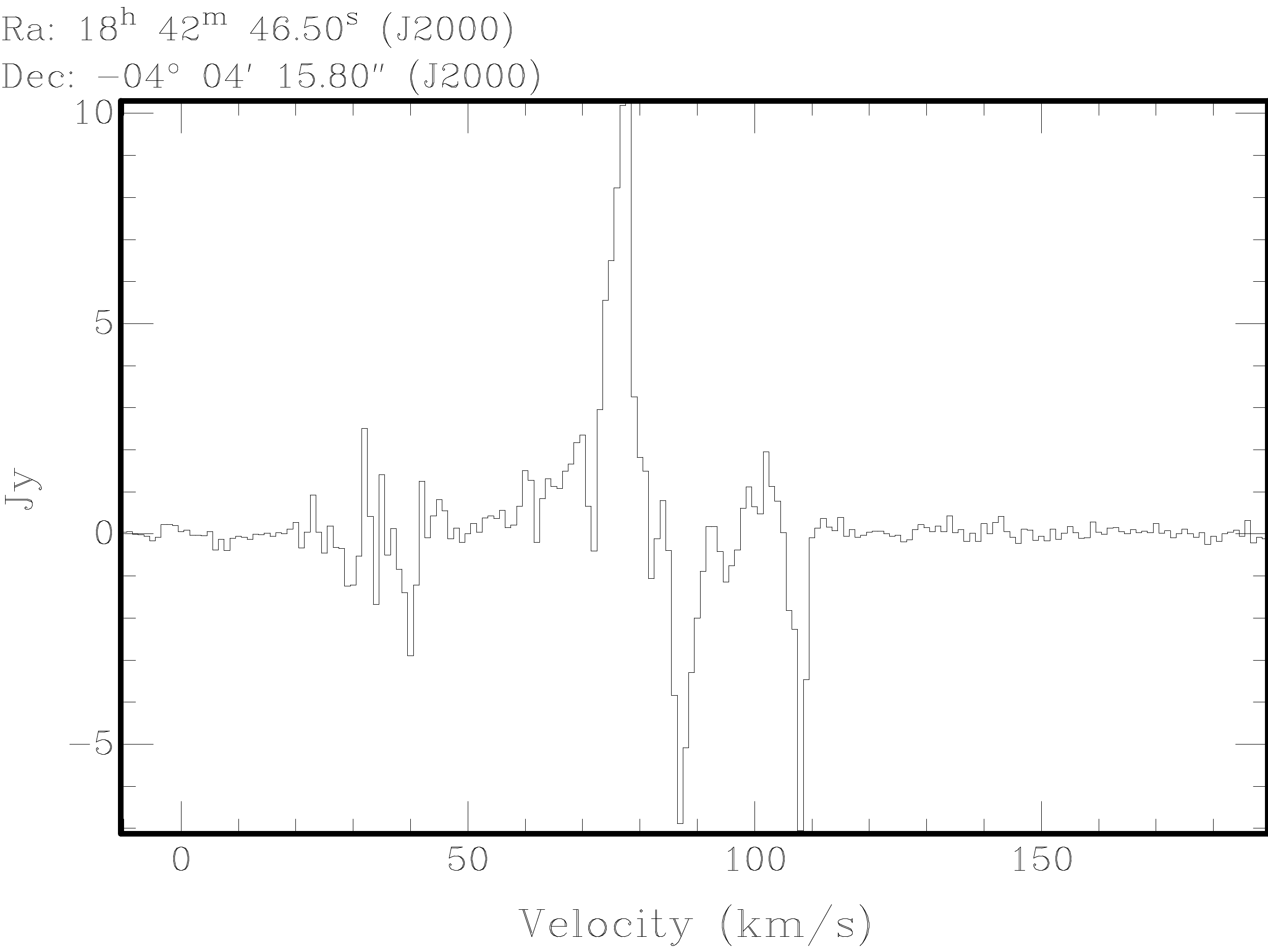}\\
\end{tabular}
\centering
\caption{CO 2-1 spectrum integrated over a 50 arcsec region around the strong
  dust continuum peak for \GO\ (top) and \GT\ (bottom). The velocities
  extend over 100\,\kms. The negative spectral line features
    likely arise from flux lost in missing spatial components with the SMA.
}
\label{appcospec}
\end{figure}

\begin{figure}
\begin{tabular}{c}
\includegraphics[height=0.2\textheight,angle=0]{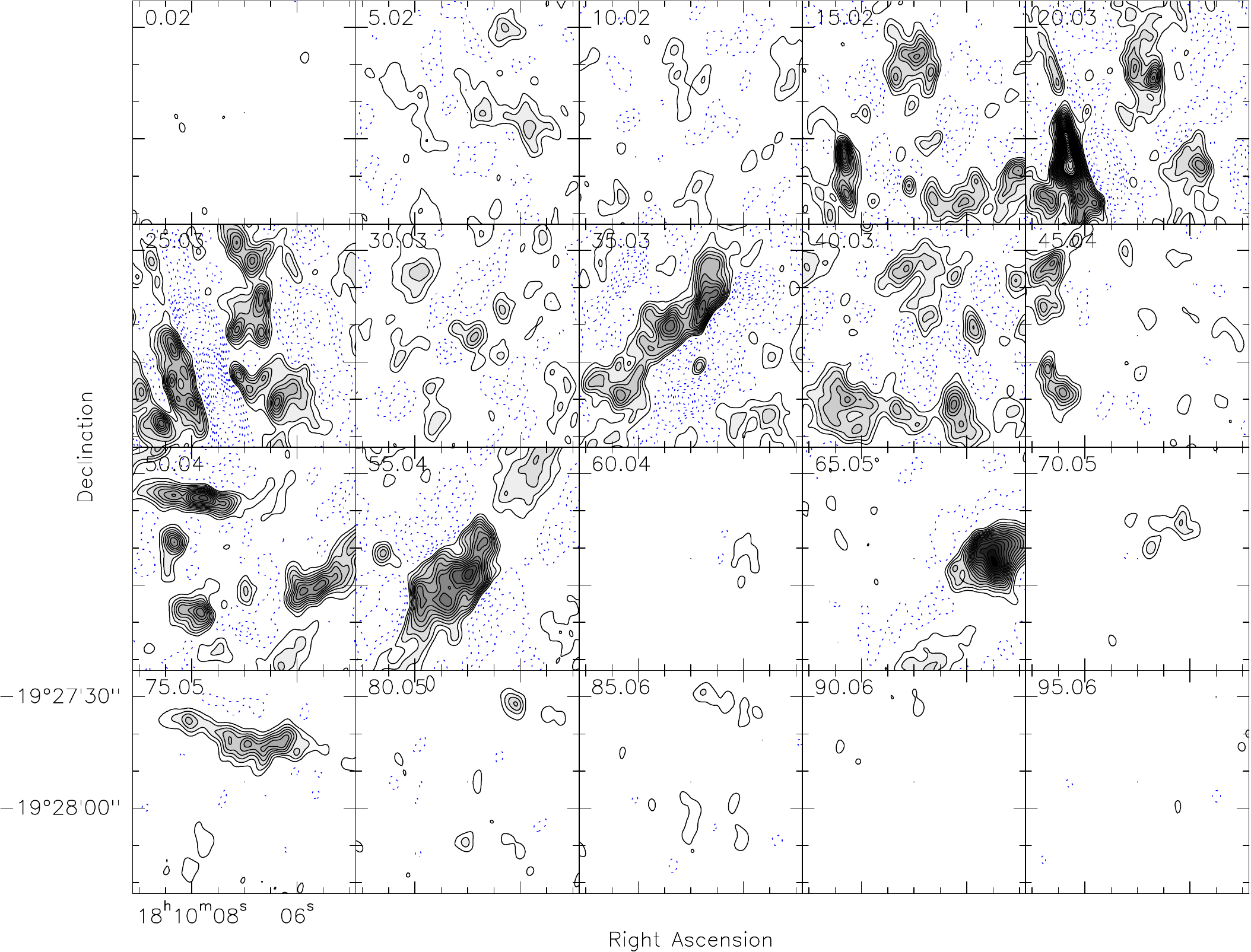}
  \\
\includegraphics[height=0.2\textheight,angle=0]{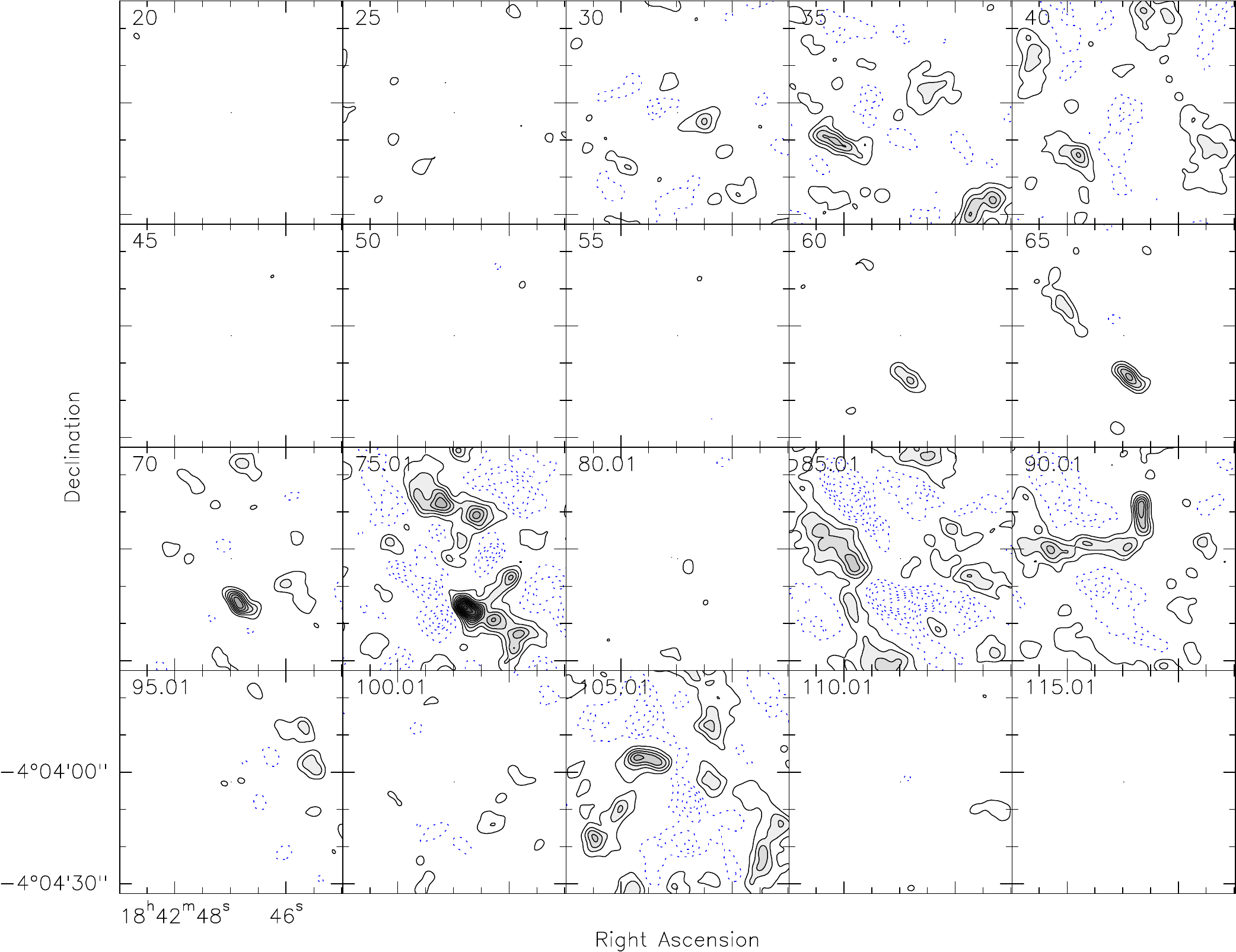}\\
\end{tabular}
\centering
\caption{CO 2-1 channel maps. These images are the result of averaging the emission over a velocity interval of 5 km/s. The central velocity is labeled on each panel. The contours start at 3\,$\sigma$ in steps of 3\,$\sigma$ . No additional $uv$ filtering has been applied to the data. The black circle in each images shows the SMA FOV at 230 GHz.}
\label{chanmap}
\end{figure}

\begin{figure*}
\begin{tabular}{c}
\includegraphics[height=0.45\textheight,angle=0]{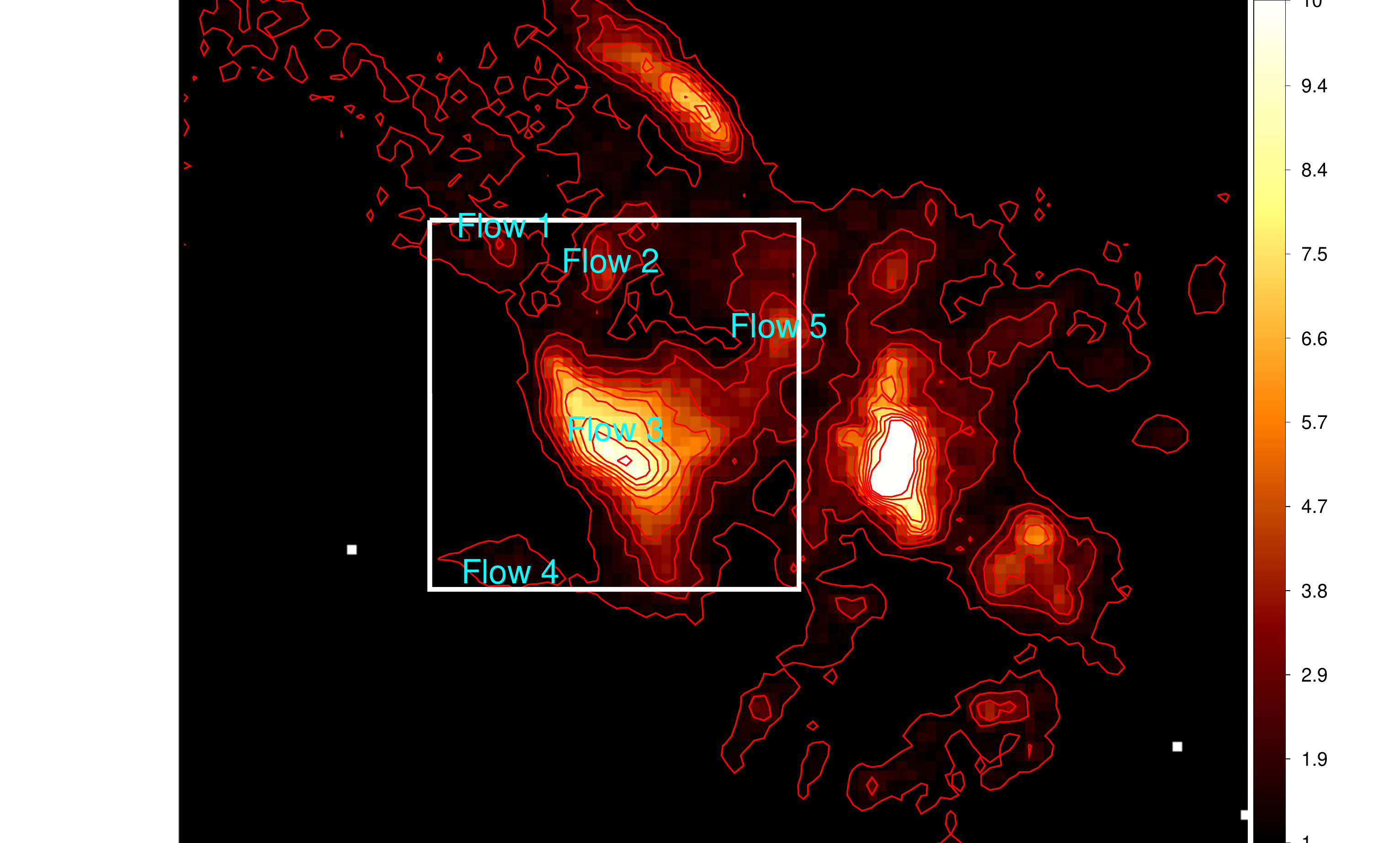}
  \\
\includegraphics[height=0.45\textheight,angle=0]{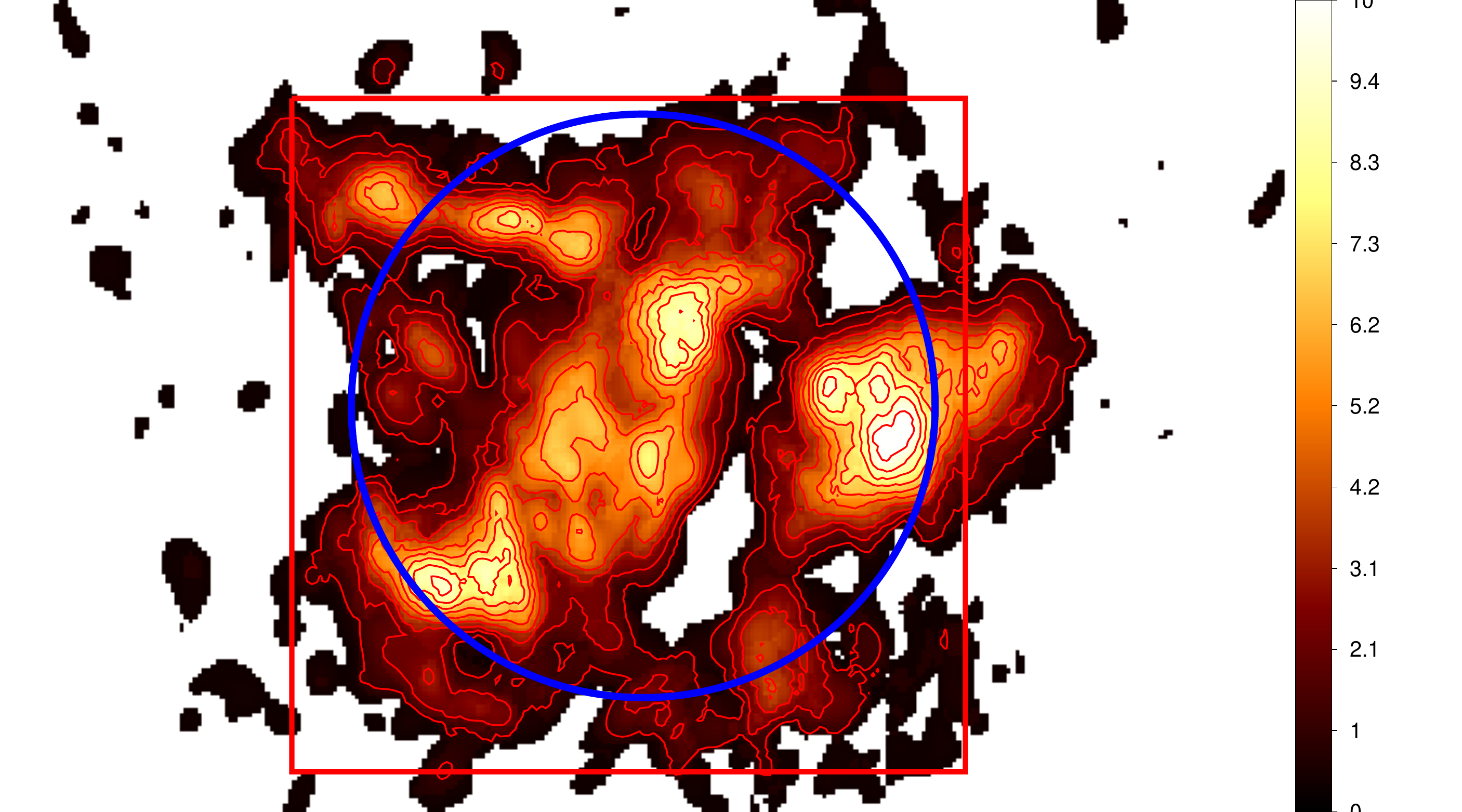}
  \\
\end{tabular}
\centering
\caption{Comparison of outflows in NGC~1333 to that observed in G10.99
  on same spatial scales. Top: CO 1-0 at 1 arcmin resolution
  integrated intensity from red-shifted (10-15 km/s)  for
  a clustered sub-region in the Perseus NGC~1333 complex scaled to the
  distance of the G10.99 region (3.6 kpc)  (colorscale and contours)
Bottom: Non-primary beam corrected SMA CO 2-1  integrated intensity
  at $\sim 3$ arcsec resolution for G10.99 for the same spatial
  scale. In both maps contours start at 1 Kkm/s up to 10 Kkm/s in steps of
  unity and the square indicates the same spatial extent of 1 arcmin. For G10.99 (bottom) the blue circle marks the SMA primary
  beam at 230 GHz.  Known outflow sources from Arce et al. 2010 within
  the NGC~1333 region are marked.}
\label{fig3}
\end{figure*}

\begin{table}[h]
\caption{Outflow upper limits for the main continuum core and integrated over the rest of the map. \label{tab:of}}
%\centering
\begin{tabular}{cccc}
feature &  Mass & Momentum & Energy \\
 &  \Msol\  & \Msol\ \kms &  \Msol\ (\kms)$^2$\\
 \hline
   & &\GO\         & \\
 main &     $>$0.01   &   $>$0.05   &   $>$0.61     \\
 rest   &     $>$0.04    &  $>$0.64    & $>$12.48   \\ 
\hline
     & &\GT\    & \\ 
main &  $>$0.03  &    $>$0.37 &     $>$5.70      \\
rest  &  $>$0.03   &   $>$0.40   &   $>$5.92      \\

\end{tabular}
\end{table}

%fluxes (peak, integratred), mass, coordinates, sizes) of the detected cores. 
%%% Table on masses
\begin{table}[h]
\caption{SMA 230 GHz continuum fluxes and masses for all dendrogram
  identified structures. \label{tab:core}}
\begin{tabular}{cccccc}
RA & Dec. & Peak flux & flux & radius & mass \\
 (J2000) & (J2000) & $\mathrm{mJy/beam}$ & $\mathrm{Jy}$ &
                                                     $\mathrm{{}^{\prime\prime}}$
                                  & \Msol\  \\
\hline
  & &  \GO\   &       \\
18:10:07.8 & -19:28:08.8 & 8.3 & 0.007 & 1.1 & 2.7\\
18:10:07.5 & -19:28:02.5 & 10.0 & 0.024 & 1.9 & 9.0 \\
18:10:07.2 & -19:27:50.6 & 7.7 & 0.004 & 0.8 & 1.5 \\
18:10:06.8 & -19:27:45.3 & 13.2 & 0.037 & 2.2 & 13.9 \\
18:10:07.0 & -19:27:36.7 & 6.6 & 0.004 & 0.8 & 1.3 \\
\hline
  & &  \GT\   &       \\

18:42:46.5 & -4:04:15.6 & 25.7 & 0.050 & 1.8 & 33.1 \\
18:42:46.9 & -4:04:07.7 & 12.7 & 0.021 & 1.5 & 14.0 \\
18:42:47.6 & -4:04:07.0 & 7.3 & 0.004 & 0.8 & 2.7 \\
18:42:47.2 & -4:03:52.6 & 7.0 & 0.013 & 1.6 & 8.7 \\

\end{tabular}
\end{table}

\end{appendix}%

\end{document}